\documentclass[11pt]{article}

\usepackage{graphicx}
\usepackage[letterpaper, margin=1in]{geometry}  
\usepackage{epstopdf}
\usepackage{amsmath,amsthm,amssymb}
\usepackage{multirow}
\usepackage{wrapfig}
\usepackage{float}
\usepackage{cite}

\usepackage{sparklines}
\usepackage{bbm}
\usepackage{xcolor}
\usepackage{setspace}
\usepackage[font=small, labelfont=bf]{caption}

\usepackage[
    colorlinks=true,
    linkcolor=blue,    % Color for internal links (ToC, Sections, Eqs)
    citecolor=blue,    % Color for bibliographical citations
    urlcolor=blue,     % Color for external URLs
    filecolor=blue     % Color for local files
]{hyperref}

\graphicspath{{./}}

\renewcommand{\vec}[1]{\boldsymbol{#1}}

\newcommand{\tens}[1]{\boldsymbol{\mathsf{#1}}}

\newcommand{\Eq}[1]{Eq.~(\ref{#1})}

\newcommand{\Fig}[1]{Fig.~\ref{#1}}

\renewcommand{\S}{\tens S}
\newcommand{\Sp}{\tens S^+}
\newcommand{\Spfull}{\tilde{\tens S}^+}
\newcommand{\mfull}{\tilde{\tens m}}
\newcommand{\mredfull}{{\mfull^{(red)}}}
\newcommand{\mm}{\tens m}
\newcommand{\M}{\tens M}

\newcommand{\Aeff}[1]{\mathbbm{A}^{(#1)}}
\newcommand{\Veff}[1]{\mathbbm{V}^{(#1)}}

\title{Improved Implementation of Approximate Full Mass Matrix Inverse Methods into Material Point Method Simulations}

\author{
   John A. Nairn \\[ -0.8ex]
    \small\emph{Oregon State University, Wood Science \&\ Engineering, 112 Richardson Hall}, \\[ -1ex] 
    \small\emph{Corvallis, OR 97330, USA}, {\tt john.nairn@oregonstate.edu}
 }

\begin{document}

\maketitle

% I think elsarticle messes this up if defined sooner
\renewcommand{\thefootnote}{\fnsymbol{footnote}}

%:Abstract
\begin{abstract}
Approximate full mass matrix methods for the material point method, known as FMPM($k$) of order $k$, can improve the calculation of grid velocities from grid momentum. It can be implemented in any MPM code by inserting a new calculation task whenever grid velocities are needed. The implementation recommended in this paper only needs these calculations once per time step just before when updating particle positions and velocities. FMPM implementation issues arise, however, when its methods are mixed with other MPM feature that rely on lumped mass calculations. Some common lumped-mass MPM features are grid-based, velocity boundary condition, multimaterial contact calculations, crack contact calculations, and imperfect interfaces. This paper first derives a revised FMPM($k$) implementation that both simplifies and clarifies the ``FMPM Loop'' that can be added to MPM codes. Next, that loop is modified to allow FMPM($k$) to work well even in simulations that need other MPM features that previously caused conflicts. Two other FMPM($k$) issues are apparent loss of stability at very higher order $k$ and inherent computational cost. These issues are discussed in an analysis of temporal stability as a function of order $k$ and in consideration of options to improve efficiency.

\vspace{9pt}
\end{abstract}

%:Keywords
%\begin{keyword}
%Material Point Method \sep Full Mass Matrix \sep Computational Mechanics
%\end{keyword}

\section{Introduction}

Approximate full mass matrix methods were developed to reduce noise and enhance stability of material point method (MPM) simulations \cite{Nairn:179,Nairn:198}. The method, denoted as FMPM($k$) for ``Full mass matrix MPM'' of order $k$, adds $k-1$ extra extrapolations within each MPM time step (or in selected time steps when using periodic FMPM($k$) \cite{Nairn:198}). The extra extrapolations can be justified either by reduction of null-space noise \cite{Nairn:179} or as an approximation to the the full mass matrix inverse that can improve calculation of nodal velocities from nodal momenta \cite{Nairn:198}. Many calculations show improved results often displaying convergence to optimal results as order $k$ increases. The approach is also amenable transport analysis where mass matrix is replaced by a ``capacity'' matrix  \cite{Nairn:2024}.

Despite documented improvements, FMPM($k$) as previously published has at least three issues:

\begin{enumerate}

\item MPM features such as grid-based boundary conditions, contact \cite{Bardenhagen:2001,Nairn:181,Nairn:190}, crack contact \cite{CRAMP,Nairn:2025}, imperfect interfaces \cite{Nairn:122,Nairn:146}, and penalty methods \cite{GUILKEY2021} rely on nodal calculations derived using a lumped mass matrix. Using prior FMPM($k$) in conjunction with such features had conflicts than that could degrade results as $k$ increased (\emph{e.g.}, boundary condition example in Ref.~\cite{Nairn:198}). Although Refs.~\cite{Nairn:198} and \cite{Nairn:190} proposed methods to deal with boundary conditions and contact, further improvements are needed.

\item Replacing lumped mass matrix with an approximate full mass matrix changes the dynamics of MPM. This change can affect FMPM($k$) stability (\emph{e.g.}, impose a need for smaller time steps) or could potentially become ill conditioned (\emph{i.e.}, full mass matrix approach singularity). For FMPM($k$) to be used with confidence, the stability of the method needs clarification and options to improve stability might be needed.

\item FMPM($k$) has unavoidable computational cost as $k$ increases meaning it can be impractical to reach ideal results. Fortunately low $k$ is often enough. Any method that can improve efficiency would be desirable.

\end{enumerate}

This paper presents a revised FMPM($k$) method that fully or partially resolves all these issues. The process first revises FMPM($k$) calculations by converting them to an incremental approach. Each increment finds the change in grid velocities between using FMPM($k$) and using FMPM($k-1$). By implementing lumped-mass matrix methods, such as contact or grid-based boundary conditions, on each increment, this revised approach resolves all conflicts of FMPM($k$) with those features (issue 1).

Next, a freely-vibrating rod problem was found to reveal temporal stability of FMPM($k$) as function of order $k$ (issue 2). The time step needed for stable FMPM($k$) calculations decreases as order increases. Fortunately it plateaus for high order meaning FMPM($k$) can be stable for any order without requiring impractically short time steps. Two options for improving stability are to blend the full mass matrix with a lumped mass matrix or to switch to periodic FMPM($k$). These options offer some benefits, but have some drawbacks.

Finally by evaluating the magnitude of each incremental change, FMPM($k)$ can be checked for convergence. A new option, therefore, is to increase efficiency by dynamically varying order $k$ during a simulation (issue 3). One example shows potential but finding metrics to check for convergence remains a challenge.

The algorithms in this paper were all implemented and tested in {\tt OSParticulas} code \cite{Nairn:Other2}. They are also publicly available in {\tt NairnMPM} version 19 or newer \cite{Nairn:Git}. Both these codes of thus examples of fully implementing FMPM($k$) calculations in MPM code.

\section{Revised FMPM($k$) Implementation}

\subsection{MPM Extrapolations and Time Step \label{MPMTimeStep}}

This section provides some MPM background, describes various MPM shape functions, outlines the calculations done in each MPM time step, and defines this paper's nomenclature. A more elaborate discussion of these concepts is available in Ref.~\cite{Nairn:198}.
MPM carries evolving simulation results on a collection of particles (or material points) but solves equations of motion on a collection of nodes in a background grid \cite{Sulsky:1994,Sulsky:1995,Bardenhagen:2004}. Particle quantities are denoted here with upper case vectors (\emph{e.g.}, $\vec V$ for particle velocities) of length $N$ where $N$ is number of particles, while nodal quantities are denoted with lower case vectors  (\emph{e.g.}, $\vec v$ for grid velocities) of length $n$ where $n$ is number of nodes. Vectors with subscripts (\emph{e.g.}, $\vec V_p$ or $\vec v_i$) indicate property for one particle or one grid node. 

Each time step maps information from the particles to the grid, updates grid values, and then maps updated information back to the particles to update their state. Linear velocity mappings can be are written as:
\[
      \vec V = \S\vec v \qquad {\rm and} \qquad  \vec v = \Sp\vec V
\]
where $\S$ is an $N\times n$ matrix of shape functions that interpolates grid values to particle locations. Any shape functions can be used although generalized interpolation MPM (or GIMP) \cite{Bardenhagen:2004} are preferred (and GIMP includes CPDI \cite{Sadeghirad:2011,Nairn:159} methods). All GIMP methods integrate over particle domain using standard grid shape functions, $\vec N_i(\vec X_p)$, which can use either linear or quadratic spline functions \cite{Nairn:198,Steffen:2008}. The reverse mapping from particles to the grid, written above as an $\Sp$ shape function matrix, would ideally invert $\S$, but that is not possible with a large, non-square $\S$. Another choice would to find the Moore-Penrose or pseudo-inverse of $\S$ \cite{Penrose:1955aa}, but that calculation is precluded by large size of $\S$. Instead, MPM uses a reverse mapping derived by least squares minimization of mass-weighted, velocity extrapolation error \cite{Sulsky:1994}:
\[
      \vec v = \mfull^{-1}\S^T\M\vec V \qquad {\rm with} \qquad \mfull = \S^T\M\S \qquad \left(i.e,\ (\mfull)_{ij} = \sum_p M_pS^T_{ip}S_{pj}\right)
\]
Here $\mfull$ is the $n\times n$, symmetric full mass matrix and $\M$ is an $N\times N$ diagonal matrix with particle mass $M_p$ on the $p^{th}$ diagonal. The resulting reverse mapping of particle velocity to the grid becomes:
\[
      \vec v = \Spfull\vec V \quad \implies \quad \Spfull = \mfull^{-1}\S^T\M
\]
Even this alternative reverse mapping needs a large, full mass matrix inverse. Standard MPM handles this issue by switching to an approximate, diagonal, lumped mass matrix, $\mm$, where sums of $\mfull$ columns are on the diagonals resulting \cite{Sulsky:1994,Sulsky:1995}:
\[
     \mm = {\rm diag}(\tens S^T\vec M)
     \qquad {\rm and} \qquad 
      \Sp = \mm^{-1}\S^T\M \qquad {\rm or} \qquad S_{ip}^+ = \frac{M_pS_{pi}}{m_i}
\]
where $\vec M$ is a vector of particle masses (\emph{i.e.}, $\M = {\rm diag}(\vec M)$).
Note that $\mfull$ and $ \Spfull$ with a tilde denote full mass matrix while $\mm$ and $\Sp$ without a tilde denote lumped mass matrix.

Although finding full mass matrix inverse is impractical, Nairn and Hammerquist \cite{Nairn:198} derives an asymptotic expansion to approximate the inverse. First, the full mass matrix can be expressed in terms of lumped-mass terms as follows:
\begin{equation}
  \mfull = \mm\Sp\S = \mm\bigl(\tens I_n - (\tens I_n - \Sp\S)\bigr) = \mm\bigl(\tens I_n -\tens A\bigr)      \label{massfullform}
\end{equation}
where $\tens I_n$ is an $n\times n$ identity matrix and $\tens A = \tens I_n - \Sp\S$. Although $\Sp$ is not an inverse to $\S$, it was selected to approximate an inverse. When that approximation is reasonable, $\tens A$ should be small and the full mass matrix inverse can be expanded in a converging Taylor series:
\begin{equation}
    \mfull^{-1} = (\tens I_n -\tens A\bigr)^{-1}\tens m^{-1} = (\tens I_n + \tens A + \tens A^2 + \tens A^3 +\cdots)\mm^{-1}
    \label{mkfull}
\end{equation}
Nairn and Hammerquist \cite{Nairn:198} defines $\mfull_k^{-1}$ as the full mass matrix inverse expanded  to $k$ terms (last term is $\tens A^{k-1}$) and proposes FMPM($k$) as  a new approach to MPM that replaces $\mm^{-1}$ with $\mfull_k^{-1}$ whenever beneficial. 

\begin{figure}
\centering
\includegraphics[width=\linewidth]{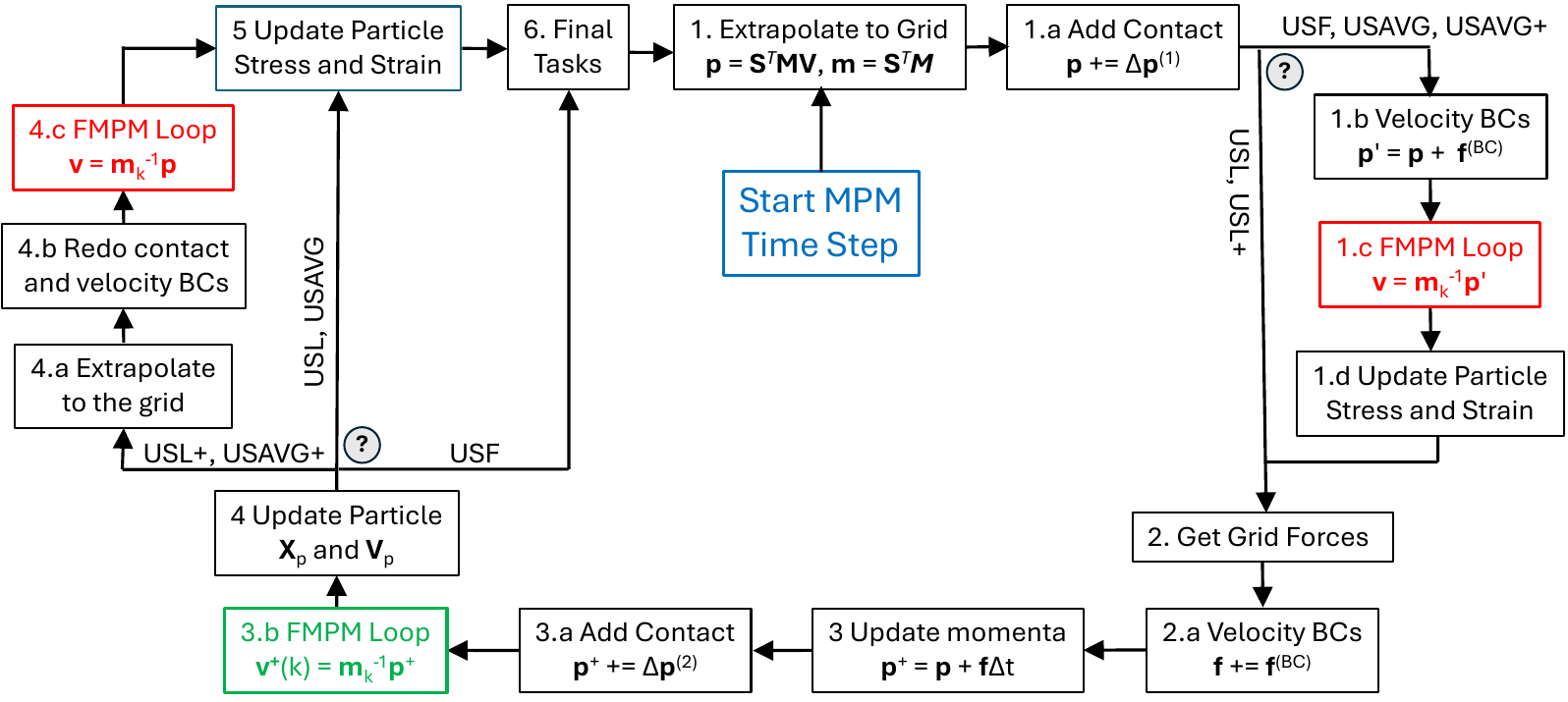}
\caption{Flow chart for an MPM time step. The branches at the two ``?'' decision points refer to calculations using USF, USL, USL+, USAVG, and USAVG+ methods for updating particle stresses and strains. All FMPM($k$) calculations are in tasks 1.c, 3.b, and 4.c. When using the recommended USL method, the only FMPM($k$) task in each time step is 3.b.}
\label{MPMFlowChart}
\end{figure}

A flow chart for each MPM time step, with modifications for contact, boundary conditions, various update schemes, and FMPM($k$) is given in \Fig{MPMFlowChart} (this organization is based on {\tt OSParticulas} \cite{Nairn:Other2} or {\tt NairnMPM} \cite{Nairn:Git} MPM codes). The six cores tasks for the recommended USL method, with more details on each step found in Ref.~\cite{Nairn:198}, are as follow (the meaning of USL and several alternative methods are explained below this core task list):

\begin{enumerate}

\item \emph{Extrapolate to Grid}: Extrapolate particle momenta ($M_p\vec V_p$) and mass ($M_p$) to grid momenta ($\vec p = \S^T\M\vec V$) and nodal masses ($\vec m = \S^T\vec M$ with lumped mass matrix $\mm = {\rm diag}(\vec m)$). When modeling multimaterial or crack contact, task 1.a adds momentum change $\Delta\vec p^{(1)}$ due to contact effects \cite{Nairn:190}.

\item \emph{Get Grid Forces}: Extrapolate Kirchoff stress ($\vec \tau_p$) and body force ($\vec B_p$) on the particles to get grid forces using $\vec f =  -\tens G^T\tens \Omega \tens \tau + \S^T\M\vec B$ where $\tens G^T$ is a matrix of shape function gradient vectors and $\tens \Omega$ is a diagonal matrix with undeformed, initial particle volume $V^{(0)}_p$ on the $p^{th}$ diagonal \cite{Nairn:190}. For simulations with grid velocity conditions, task 2.a add the reaction forces $\vec f^{(BC)}$ needed to satisfy those conditions to the grid forces.

\item \emph{Update Grid Momenta}: Update momenta on the grid using total grid forces found in previous task as $\vec p^+ = \vec p + \vec f\Delta t$. When modeling multimaterial or crack contact, task 3.a adds momentum change $\Delta\vec p^{(2)}$ due to contact effects (note that Ref.~\cite{Nairn:190} explains why contact calculations are needed twice each time step --- in task 1 and task 3).

\vskip 3pt
\emph{Task 3.b}: When using FMPM($k$) find updated grid velocities $\vec v^+(k) = \mfull_k^{-1}\vec p^+$. When using lumped mass matrix methods, the FMPM loop is skipped and replaced with finding lumped-mass grid velocities $\vec v^+(1) = \mm^{-1}\vec p^+$ and grid accelerations $\vec a = \mm^{-1}\vec f$. Note that lumped grid velocities are equivalent to FMPM(1) grid velocities.

\item \emph{Update Particle Position and Velocity}:  Use grid velocities and accelerations found in task 3.b to update particle position $\vec X_p$ and velocity $\vec V_p$ using FMPM($k$) methods \cite{Nairn:198} or FLIP methods \cite{FLIP_0} (see below).

\item \emph{Particle Stress and Strain}: Extrapolate grid velocities to particles using gradient shape functions, $\tens G$, to find particle velocity gradients, $\nabla\vec V_p^T = \tens G \vec v^+(k)$, on the particles. Lumped mass matrix methods are special case of this calculation using $\vec v^+(1)$. Once $\nabla\vec V_p^T$ are found, calculate increment in deformation gradient using $d\tens F_p = \exp{(\nabla\vec V_p\Delta t)}$ and use it to model any chosen constitutive law.

\item \emph{Final Tasks}: Implement any needed tasks such as detecting when particles move between elements or between patches used in parallel computation, moving cracks, or any added custom task calculations.

\end{enumerate}

The above steps are core to MPM calculations. The only change needed to implement FMPM($k$) for USL method is calculation $\vec v^+(k)$ in task 3.b. This paper introduces new methods to keep contact and velocity boundary conditions compatible with FMPM($k$) and those methods can all be coded within a revised FMPM($k$) loop in task 3.b.

Note that task 5 updates particle stress and strain after updating particle position and velocity and is labeled USL for ``Update Strains Last.'' Other MPM options are USF to update strains first using tasks 1.b, 1.c, and 1.d \cite{Bardenhagen:2002}, USL+ to re-extrapolate particle state to the grid in tasks 4.a, 4.b, and 4.c  before updating in task 5 \cite{Zhou:1998}, or use both USF and USL (or USL+) and average them on each time step \cite{CRAMP} denoted as USAVG (or USAVG+). The MPM time step for each of these update methods follows the appropriate branch at each decision point ``?'' in Table~\ref{MPMFlowChart}.

The relative time for FMPM($k$) compared to conventional MPM is \cite{Nairn:198}
\begin{equation}
      {\rm Relative\ Time} = 1 + \phi(k-1)F.  \label{compcost}
\end{equation}
where $F=C_x/C_0$ is the fraction of a time step spent finding grid velocities in tasks 1.c, 3.b, or 4.c (\emph{i.e.}, the colored blocks in \Fig{MPMFlowChart})  and $\phi$ is the number those tasks needed on each time step.  When using FMPM($k$), the USL method is preferred for both efficiency and theoretical reasons. It is most efficient because $\phi=1$. In contrast, USF has $\phi=2$ (tasks 1.c and 3.b), USL+ has $\phi=2$ (tasks 3.b and 4.c), USAVG has $\phi=2$ (tasks 1.c and 3.b), and USAVG+ has $\phi=3$ (tasks 1.c, 3.b, and 4.c) \cite{Nairn:198}. The theoretical reason to prefer USL is that it is the only method that updates particle position and velocity with the same grid velocities used to update particle stress and strain. Those grid velocities are found in task 3.b and then used in both tasks 4 and 5. Some prior results suggested methods other than USL might have better energy conservation \cite{CRAMP} especially when modeling contact. Those findings are suggested below as illusory. 

\subsection{Approximate Full Mass Matrix Implementation or the ``FMPM Loop''}

FMPM($k$) implementation in Ref.~\cite{Nairn:198} summed expansions of each $\tens A^j$ term and recast the result as:
\begin{equation}
      \vec v^+(k) = \mfull_k^{-1}\vec p^+ = \left(\sum_{\ell=1}^{k} (-1)^{\ell+1}{k \choose \ell}\left(\Sp\S\right)^{\ell-1}\right)\mm^{-1}\vec p^+   = 
      \sum_{\ell=1}^k (-1)^{\ell+1} \vec v_\ell^*         \label{mfullOld}
\end{equation}
where
\[
      \vec v^*_\ell =  {k\choose \ell}(\tens S^{+}\tens S)^{\ell-1}\mm^{-1}\vec p^+ = \frac{k+1-\ell}{\ell}\tens S^{+}\tens S\vec v^{*}_{\ell-1} 
      \quad {\rm and} \quad  \vec v_1^* = k \mm^{-1}\vec p^+ 
\]
The implementation needed only a single subroutine to calculate $\vec v^*_\ell$ from $\vec v^*_{\ell-1}$ that was seeded with $\vec v^*_1$ and then called recursively $k-1$ times. These $k-1$ recursions is the basis for the computation cost in \Eq{compcost} with $F$ being the fraction of time spent in each recursion relative to the entire MPM time step time.

This paper adopts a revised implementation. Rather than expanding $\tens A^j$ terms, the better approach is to incrementally add $\mfull_k^{-1}$ terms in \Eq{mkfull} leading to:
\begin{equation}
      \vec v^+(k) = \mfull_k^{-1}\vec p^+ = \left(\sum_{\ell=1}^{k} \tens A^{\ell-1}\right)\mm^{-1}\vec p^+   = 
      \sum_{\ell=1}^k \Delta \vec v(\ell)   \label{mfullEq}
\end{equation}
where $\Delta \vec v(\ell) = \vec v^+(\ell) - \vec v^+(\ell-1)$ is the difference in finding velocity by FMPM($\ell$) and FMPM($\ell-1$). The revised recursive relation is
\begin{align}
     \Delta \vec v(\ell) & =  \tens A^{\ell-1}\mm^{-1}\vec p^+ = \tens A\Delta \vec v(\ell-1) = \Delta\vec v(\ell-1) - \tens S^{+}\tens S\Delta\vec v(\ell-1) \label{recursion} \\
      & \quad {\rm starting\ with} \quad  \Delta \vec v(1) = \vec v^+(1)=\mm^{-1}\vec p^+  \nonumber
\end{align}
A pseudo-code algorithm for an ``FMPM Loop'' to find $\vec v^+(k)$ is given in Table \ref{FMPMcode}. The revised method is a minor change. Both prior and revised methods find $\vec v_{next}$ from $\tens S^{+}\tens S$ times a velocity --- $\vec v^{*}_{\ell-1}$ in prior method and $\Delta\vec v(\ell-1)$ in revised one (which is $\vec v_{prev}$ in pseudo-code). The prior method scales that result by $(k+1-\ell)/\ell$ while the revised method leaves it unscaled and subtracts the result from the previous velocity increment $\vec v_{prev}$. The two methods are mathematically identical, but the revised method is amenable to changes that fix grid-based velocity boundary conditions and contact calculations (see lines labels (1) and (2)). Because the revised recursion relation is now independent of $k$ (which appeared in scaling term of prior method), the new method has potential to implement dynamic methods that adapt order $k$ during a simulation (\emph{i.e.}, continue until $\Delta \vec v(\ell)$ is sufficiently small in line labeled (3)).

\begin{table}
\caption{Full ``FMPM Loop'' for revised FMPM($k$) calculation of $\vec v^+(k)$. Steps indicated as (1), (2), and (3), are extra calculations made possible by this revised method to improve grid boundary conditions, contact calculations, and support a dynamic FMPM($k$) option.}   \label{FMPMcode}
\rule{\textwidth}{0.4pt}

{
\tt\bf\setlength{\parindent}{2em}
let  $\vec v^+(1) = \vec v^* = \vec v_{prev}  \leftarrow \mm^{-1}\vec p^+$ on all active nodes

for each $\ell$ in $2$ to $k$ do

\hspace{.2in}let $\vec v_{next} \leftarrow 0$ on all active nodes

\hspace{.2in}for each particle $p$ do

\hspace{.4in}find $\vartheta_p$ nodes with non-zero shape functions ($S_{pi}$)

\hspace{.4in}for each node $i$ in $\vartheta_p$ do

\hspace{.6in}for each node $j$ in $\vartheta_p$ do

\hspace{.8in}let $(\vec v_{next})_i \leftarrow (\vec v_{next})_i + \frac{M_p S_{pi} S_{pj}}{m_i} (\vec v_{prev})_j$

\hspace{.2in}let $ \vec v_{prev} \leftarrow \vec v_{prev} - \vec v_{next}$
          \hspace{.5in}{\rm /* \emph{$\vec v_{next}=\tens S^+\tens S\vec v_{prev}$ so new $\vec v_{prev}=\tens A\vec v_{prev}$} */}

\hspace{.2in}set controlled grid velocities in $\vec v_{prev}$ to zero (1)

\hspace{.2in}adjust $\vec v_{prev}$ to remain consistent with contact laws (2)

\hspace{.2in}{\rm /* \emph{comment: corrected $\vec v_{prev}$ is now equal to $\Delta\vec v(\ell)$} */}

\hspace{.2in}let $\vec v^* \leftarrow \vec v^*  + \vec v_{prev}$
        \hspace{1in}{\rm /* \emph{$\vec v^*$ is evolving $\vec v^+(\ell)$} */}

\hspace{.2in}if $ \| \vec v_{prev} \|$ is sufficiently small, exit as converged with FMPM($\ell$) (3)

}
\rule{\textwidth}{0.4pt}
\label{FMPMCode}
\end{table}

\subsection{Particle Updates}

Generalized updates for particle velocity and position can be written as:
\begin{equation}
         \vec V^{(n+1)} =  \vec V^{(n)}+ \Aeff{n}\Delta t 
         \quad{\rm and}\quad
         \vec X^{(n+1)}  =\vec X^{(n)}+ \Veff{n}\Delta t + \alpha\Aeff{n}(\Delta t)^2 
         \label{GenericUpdates}
 \end{equation}
 where $\Aeff{n}$ and $\Veff{n}$ are effective acceleration and velocity implied by any proposed update method and $\alpha$ is a time-integration parameter. Any proposed update methods can be cast in terms of $\Aeff{n}$, $\Veff{n}$, and $\alpha$. For example, standard updates in the Full Lagrangian Implicit Particle, or FLIP method,\footnote{\label{*}Many codes use $\alpha=1$ to update using final grid velocity extrapolated to the particles; choosing $\alpha=1/2$ uses average velocity during the time step extrapolated to the particles instead of final velocity.} is \cite{Sulsky:1994,FLIP_0}:
\begin{equation}
         \vec V^{(n+1)} = \vec V^{(n)}+ \tens S\mm^{-1}\vec f\Delta t  \quad {\rm and} \quad
         \vec X^{(n+1)} = \vec X^{(n)} + \tens S\vec v^+(1)\Delta t  
              - (1-\alpha)\tens S\mm^{-1}\vec f(\Delta t)^2     \label{FLIPUpdates}
\end{equation}
where $\vec v^+(1)$ and  $\mm^{-1}\vec f = \vec a$ are updated grid velocity and acceleration calculated using a lumped mass matrix. Substituting into \Eq{GenericUpdates} leads to:
\[
         \Aeff{n}=  \tens S\mm^{-1}\vec f   \quad {\rm and} \quad
         \Veff{n}  =   \tens S\vec v^+(1) -  \tens S\mm^{-1}\vec f \Delta t = \tens S\vec v(1)
\]
The FLIP concept is that particle velocity should increment by acceleration extrapolated to the particle rather then by extrapolating a new velocity. The FLIP approach has low dissipation, but is prone to null-space noise developing in the particle values.

Rather than extrapolate acceleration, FMPM($k$) replaces particle velocities by extrapolating velocity found using $\mfull_k^{-1}$. Extending the FMPM($k$) update in Ref.~\cite{Nairn:198}, which used $\alpha=1/2$, to allow any $\alpha$, the velocity and position updates become:
\[
     \vec V^{(n+1)} = \S\vec v^+(k) \quad {\rm and} \quad
     \vec X^{(n+1)} = \vec X^{(n)} + \left(\alpha\vec V^{(n+1)}+(1-\alpha)\vec V^{(n)}\right)\Delta t
\]
Substituting into \Eq{GenericUpdates} leads to:
\begin{equation}
       \Aeff{n}  = \frac{\S\vec v^+(k)-\vec V^{(n)}}{\Delta t} = \tens S\mfull_k^{-1}\vec f + \frac{\S\vec v(k) - \vec V^{(n)}}{\Delta t} 
       \quad {\rm and} \quad \Veff{n} = \vec V^{(n)}    \label{FMPMeff}
\end{equation}
This FMPM($k$) update is analogous to Particle In Cell, or PIC, style methods that replaces particle velocity on each time step with new results extrapolated from the grid. An advantage of PIC-style updates is that they remove null-space noise that can grow on the particles \cite{Nairn:179} when using FLIP. A disadvantage is energy dissipation. For example, FMPM(1), which a lumped-mass PIC method, dissipates too much energy in most problems. Switching to FMPM($k$) (and its precursor denoted XPIC($k$) \cite{Nairn:179}) greatly reduces dissipation. With sufficiently high order, FMPM($k$) both removes null-space noise and dissipates less energy than FLIP methods \cite{Nairn:198}. Although the first $\Aeff{n}$ in \Eq{FMPMeff} is the one used in coding, the second form casts FMPM($k$) as revised FLIP update. Compared to FLIP velocity update, FMPM($k$) finds acceleration using an approximate full mass matrix inverse ($\vec a(k)=\mfull_k^{-1}\vec f$) rather the lumped mass matrix and adds a ``damping'' term. The damping term is proportional to the difference between velocity extrapolated to the grid and back ($\S\vec v(k)$) and the original particle velocity ($\vec V^{(n)}$). This damping term selectively dampens null-space noise and should decrease in magnitude as $k$ increases. 

Because $\Veff{n}$ and $\Aeff{n}$ are assumed constant during one time step, the average Lagrangian (or material) velocities for the particles can can be found from either generic position or velocity update in \Eq{GenericUpdates} as:
\[
     \frac{\Delta\vec X}{\Delta t}  = \Veff{n} + \alpha\Aeff{n}\Delta t  \quad {\rm and} \quad \left\langle \vec V\right\rangle = \vec V^{(n)} + \frac{1}{2}\Aeff{n}\Delta t
\]
Ideally, these two calculations of the same quantity would match, but in general they differ by
\[
           \frac{\Delta\vec X}{\Delta t} -\left\langle \vec V\right\rangle = \left\{\begin{array}{ll}
           \tens S\vec v(1) -\vec V^{(n)} + (\alpha-\frac{1}{2}) \tens S\mm^{-1}\vec f\Delta t & {\rm FLIP} \\
           (\alpha-\frac{1}{2})\bigl(\S\vec v^+(k)-\vec V^{(n)}\bigr) & {\rm FMPM}(k)
           \end{array}\right.
\]
These two Lagrangian velocities always differ for FLIP, which implies an inconsistency between FLIP position and velocity updates. When using FMPM($k$), the particle update inconsistency can be eliminated by choosing $\alpha=1/2$.

\section{Results and Discussion}  \label{RandD}

Each subsection below modifies the revised FMPM($k$) methods to resolve prior issues seen in FMPM($k$) calculations.

\subsection{Grid Velocity Boundary Conditions}

A common MPM boundary condition method is to impose velocities on grid nodes denoted as ``velocity BCs.'' Zero-velocity BCs model a barrier while non-zero velocity BCs can pull or push objects. For modeling large-displacements with boundary conditions, velocity BCs need to translate through the grid. 

The goal of imposing velocity BCs is to change the momentum update (see step 3 in Section \ref{MPMTimeStep}) to $\vec p^{+} = \vec p + (\vec f+\vec f^{(BC)})\Delta t$  where $\vec f^{(BC)}$ are reaction forces needed for $\vec v^{+}(k) = \mfull_k^{-1}\vec p^{+}$ to satisfy boundary conditions on nodes with BCs. Allowing each node $i$ to have one or more velocity BCs, that set velocity in direction $\hat{\vec n}_i^{(b)}$ to $v_i^{(b)}$ (for BC number $b$),  the momentum change, $\Delta\vec p = \vec f^{(BC)}\Delta t$, induced by BCs should lead to nodal velocities on BC nodes of:
\begin{equation}
          \bigl(\mfull_k^{-1}\Delta\vec p\bigr)_i = \Delta\vec v_i = 
              \sum_b \bigl(v_i^{(b)} - \vec v_i^+(k) \cdot \hat{\vec n}_i^{(b)}\bigr) \hat{\vec n}_i^{(b)}             \label{CoupledBCs}
\end{equation}
The net result of this calculation is to replace the velocity in the $\hat{\vec n}_i^{(b)}$ direction with the desired BC velocity $v_i^{(b)}\hat{\vec n}_i^{(b)}$. Each node can superpose any number of BCs provided all $\hat{\vec n}_i^{(b)}$ are perpendicular or parallel to each other.

Using an approximate full mass matrix inverse causes coupling between BC nodes and non-BC nodes. As a result, we would need $n$ equations to solve for the $n$ values of $\Delta\vec v_i$ but \Eq{CoupledBCs} defines much fewer than $n$ conditions. Recognizing these conflicts between FMPM($k$) and velocity BCs, Ref.~\cite{Nairn:198} proposed three approximate BC methods:

\begin{enumerate}

\item \emph{Lumped Method}: the simplest approach is to ignore the conflicts and use lumped mass matrix BC methods in step 2.a. When using lumped mass methods (or when using FMPM(1)), \Eq{CoupledBCs} simplifies such that momentum change on each BC node can be independently solved as
\[
       \Delta\vec v_i = \frac{ \Delta\vec p_i}{m_i} = 
              \sum_b \bigl(v_i^{(b)} - \vec v_i^+(1) \cdot \hat{\vec n}_i^{(b)}\bigr) \hat{\vec n}_i^{(b)} 
\]

\item \emph{Velocity Method}: this approach ignored BCs (\emph{i.e.}, skipped step 2.a) until after the FMPM($k$) calculations in step 3.b and then simply changed $\vec v^+(k)$ to satisfy BC velocity.

\item \emph{Combined Method}: this approach combined the above two methods; \emph{i.e.}, calculate lumped mass matrix forces in step 2.a and then adjusted $\vec v^+(k)$ to satisfy BC velocity after step 3.b.

\end{enumerate}
Both lumped and velocity methods were poor for any FMPM($k>1$). The lumped method is poor because the FMPM($k$) calculations that come after finding $\vec f^{(BC)}$ can change velocities on BC nodes to no longer satisfy the BCs. Although velocity method exactly satisfies velocity BCs, the effects of velocity BCs are felt only on nodes with BCs, which is contrary to expectations for full mass matrix calculations. The combined method worked better, but results eventually degraded as $k$ increased.

This paper defines a new approach that was made possible by revision of FMPM($k$) based on incremental velocity changes $\Delta\vec v(\ell)$. The new approach is to impose velocity BC constraints on each velocity increment. The FMPM($k$) velocity calculation then becomes
\[
       \vec v^+(k) = \sum_{\ell=1}^k \left(\Delta\vec v(\ell)^{(0)} + \Delta\vec v(\ell)^{(BC)} \right)
\]
where $\Delta\vec v(\ell)^{(0)}$ is velocity increment ignoring velocity BCs and $\Delta\vec v(\ell)^{(BC)}$ is velocity change needed to satisfy velocity BC constraints if the calculations stopped at FMPM($\ell$). The calculations start by using standard lumped-mass methods in step 2.a, which correspond to $\Delta\vec v(1)^{(0)}$ being the initial lumped-mass velocity on node $i$ and
\begin{equation}
      \Delta\vec v(1)^{(BC)} = \bigl(v_i^{(b)} - \Delta\vec v(1)^{(0)} \cdot \hat{\vec n}_i^{(b)}\bigr) \hat{\vec n}_i^{(b)} 
      \label{StartBCCalcs}
\end{equation}
where a sum of parallel or perpendicular conditions $b$ is implied. Next, evaluate final velocity in the $\hat{\vec n}_i^{(b)}$ direction:
\begin{equation}
         \vec v^+(k)\cdot\hat{\vec n}_i^{(b)} = v_i^{(b)} + \sum_{\ell=2}^k 
                \left(\Delta\vec v(\ell)^{(0)} + \Delta\vec v(\ell)^{(BC)} \right)\cdot \hat{\vec n}_i^{(b)}     \label{bcResultant}
\end{equation}
The result can be made to satisfy $\vec v^+(k)\cdot\hat{\vec n}_i^{(b)} = v_i^{(b)}$ by choosing
\[
 \Delta\vec v(\ell)^{(BC)} = - \left(\Delta\vec v(\ell)^{(0)} \cdot \hat{\vec n}_i^{(b)}\right) \hat{\vec n}_i^{(b)}
\]
such that each term in the sum is zero. Comparing to \Eq{StartBCCalcs}, the required $\Delta\vec v(\ell)^{(BC)}$ is the velocity change needed to satisfy zero additional velocity in direction $\hat{\vec n}_i^{(b)}$. The proposed, new velocity BC algorithm, with pseudocode, is:

\begin{enumerate}

\item Implement lumped mass method velocity BCs in \Eq{StartBCCalcs} such that $\vec v^{+}_i(1) \cdot\hat{\vec n}_i^{(b)} =v_i^{(b)}$ for nodes with BCs (pseudocode for this step is top half of Table~\ref{BCCode}). This step is implied in Table~\ref{FMPMCode} by the first line starting with  $\vec v^{+}(1)$.

\item After finding $\vec v_{prev} = \Delta\vec v(\ell)^{(0)}$ during pass $\ell$ in Table~\ref{FMPMCode}, set BC nodes in $\vec v_{prev}$ to be zero in BC directions (line labeled (1) in Table~\ref{FMPMCode} with pseudocode for this step in bottom half Table~\ref{BCCode}).

\end{enumerate}

\begin{table}
\caption{Implementation of velocity BCs that superposes multiple BCs on each node. The top block is done prior to FMPM($k$) loop in Table~\ref{FMPMCode} while the bottom block explains line (1) in that table. The first internal loop in each block zeros the $\hat{\vec n}_i^{(b)}$ direction. The second internal loop in the top block superposes the BC velocities. Multiple BCs on a single node must be perpendicular or parallel to each other. For parallel BCs, the first  one zeros the velocity in that BC direction. Subsequent BCs in the same direction use the evolving $\vec v_i^{+}(1)$ for which ${\vec v_i^{+}(1)} \cdot \hat{\vec n}_i^{(b)}=0$, meaning subsequent BC will correctly subtract zero.}
\rule{\textwidth}{0.4pt}

{
\tt\bf\setlength{\parindent}{2em}
{\rm /* \emph{Initial calculation of lumped-mass velocity BCs} */}

for each node $i$ with velocity BCs

\hspace{.2in}for each BC $b$ on node $i$ do

\hspace{.4in}let ${\vec v_i^{+}(1)} \mathbin{{-}{=}} \bigl({\vec v_i^{+}(1)} \cdot \hat{\vec n}_i^{(b)}\bigr) \hat{\vec n}_i^{(b)}$

\hspace{.2in}for each BC $b$ on node $i$ do

\hspace{.4in}let ${\vec v_i^{+}(1)} \mathbin{{+}{=}} v_i^{b} \hat{\vec n}_i^{(b)}$
}

\rule{\textwidth}{0.4pt}

{
\tt\bf\setlength{\parindent}{2em}
{\rm /* \emph{Set zero velocity BC in incremental velocities} */}

for each node $i$ with velocity BCs

\hspace{.2in}for each BC $b$ on node $i$ do

\hspace{.4in}let $\vec v_{prev} \mathbin{{-}{=}} \bigl(\vec v_{prev} \cdot \hat{\vec n}_i^{(b)}\bigr) \hat{\vec n}_i^{(b)}$

}
\rule{\textwidth}{0.4pt}
\label{BCCode}
\end{table}

To net effect on FMPM($k$) calculations is illustrated by including the $\Delta\vec v(\ell)^{(BC)}$ terms
\begin{align}
      \vec v^{+}(2) & = \vec v^{+}(1)  + \Delta\vec v(2)^{(0)}+\Delta\vec v(2)^{(BC)} = (\tens I+ \tens A)\vec v^{+}(1) + \Delta\vec v(2)^{(BC)}   \nonumber\\
      \vec v^{+}(3) & = \vec v^{+}(2) + \tens A(\tens A\vec v^{+}(1) + \Delta\vec v(2)^{(BC)})+\Delta\vec v(3)^{(BC)}  \nonumber \\
                   & = (\tens I+ \tens A+\tens A^2)\vec v^{+}(1) + (\tens I+\tens A) \Delta\vec v(2)^{(BC)} +\Delta\vec v(3)^{(BC)} \nonumber \\
                   & = \mfull_3^{-1}\vec p^+ + (\tens I+\tens A) \Delta\vec v(2)^{(BC)} +\Delta\vec v(3)^{(BC)} \nonumber \\
                   & \vdots \nonumber \\
      \vec v^{+}(k) & = \mfull_k^{-1}\vec p^+ + \sum_{\ell=2}^k \left(\sum_{j=0}^{k-\ell} \tens A^j \Delta\vec v(\ell)^{(BC)}\right)
              \label{genBC}
\end{align}
The first term in $\vec v^{+}(k)$ spreads lumped-mass-matrix velocity BCs to nearby nodes depending on the bandwidth of $\mfull_k^{-1}$ while the second term prevents that spread from violating any velocity BCs.

This new method for $\vec v^{+}(2)$ is identical to the combined method in Ref~\cite{Nairn:198} because that method also corrected velocities at BC nodes after finding $\vec v^{+}(2)$. It diverges for $k>2$, because the new method applies BCs to each velocity increment rather than only applying them to the final $\vec v^{+}(k)$. Because Ref~\cite{Nairn:198} reports the combined method worked for $k=2$ but degraded for $k>2$, the new method will continue to work for $k=2$ and hopefully will fix prior issues for $k>2$. 

\begin{figure}
\centering
\includegraphics[width=0.6\linewidth]{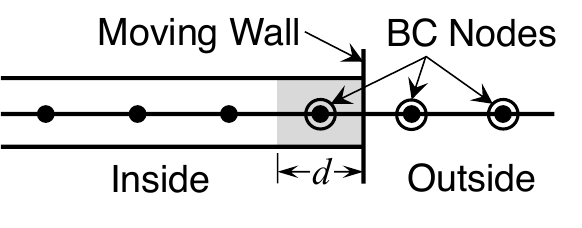}
\caption{A moving wall on the right edge of a particle domain should set velocity boundary conditions on all circled nodes. $d$ is the wall depth that determines which nodes inside the particle domain get velocity boundary conditions.}
\label{WallBC}
\end{figure}

The new velocity BC method was validated, and its improvement over the prior ``combined'' demonstrated, by repeating the manufactured solution (MMS) for a uniaxial strain problem with known, exact solution \cite{Kamojjala:2013} that was used to test moving wall conditions in Ref.~\cite{Nairn:198}. Imagine a bar pulled by a wall on the right edge of the specimen moving in the $x$ direction at velocity $v^{(b)}$ (see 1D view in \Fig{WallBC}). On each time step the wall projects velocity BCs to nearby nodes, which includes all nodes outside the object having non-zero mass, and all nodes inside the object to a specified depth $d$ from the wall. The dynamically selected nodes are assigned to velocity BC in $x$ direction of:
\[
        v_i^{(b)} = v^{(b)} + \nabla v^{(b)}(x_i - x_{wall})
\]
where $\nabla v^{(b)}$ is the spatial velocity gradient for the deformation, and $x_i$ and $x_{wall}$ are locations for node $i$ and the wall, respectively. 

New velocity BCs were validated by modeling a $20\times4$~m bar subjected to position-independent deformation gradient with $F_{xx}=1+\dot\varepsilon t$, $F_{yy}=F_{zz}=1$ and all other components zero, where $\dot\varepsilon$ is a constant strain rate. The material was modeled as a Neohookean material with Cauchy stress given by \cite{Ogden:1984,Zienkiewicz:2000}:
\begin{equation}
        \tens\sigma = \frac{\lambda}{2}\left(J-\frac{1}{J}\right)\tens I + \frac{G}{J}\left(\tens F\tens F^T- \tens I\right)     \label{NHMat}
\end{equation}
These calculations used $G=0.375$~Pa and $\lambda=0$ (or E=0.75~Pa and $\nu=0$) such that $\sigma_{yy}=\sigma_{zz}=0$ and the exact solution has zero transverse displacement. The MPM simulation used 0.25~m cells, USL update, and two material points per cell in each direction. Setting $x=0$ at the bar center, each particle was initialized with velocity $\vec V_p(0) = (\dot\varepsilon X_p,0,0)$. The deformation was controlled solely by moving wall velocity BCs on each end that set spatial velocity and gradient in the $x$ direction to
\[
      \vec v^{(b)} = \frac{\dot\varepsilon x_{wall}}{1+ \dot\varepsilon t} \quad {\rm and} \quad
      \nabla v^{(b)} = \frac{\dot\varepsilon}{1+ \dot\varepsilon t} \quad \implies \quad
      \vec v_i^{(b)} = \frac{\dot\varepsilon x_i}{1+ \dot\varepsilon t} 
\]
Because $\lambda=0$, no velocity BCs were needed to constrain transverse deflection. Because this problem has zero acceleration, the particle velocities should remain constant (with their initial values) and the axial stress should be
\[
   \sigma_{xx}  =  2G\frac{\left(2+ \dot\varepsilon t\right)\dot\varepsilon t}{2\left(1 + \dot\varepsilon t\right)}
\]
The simulations elongated the specimen by 20\% such that the moving walls crossed eight grid cells during the simulation. The plotted velocity error (in percent) is an average of RMS velocity error over the full simulation normalized to the end velocity ($V_{end}=10\dot\varepsilon$):
\[
     \left\langle{\rm Velocity\ Error}\right\rangle\ (\%) = \frac{100}{N_sV_{end}} \sqrt{\frac{1}{N}\sum_p \|\vec V_p(t)-\vec V_p(0)\|^2}
\]
where $N_s$ is the number of averaged time steps.

\begin{figure}
\centering
\includegraphics[width=0.45\linewidth]{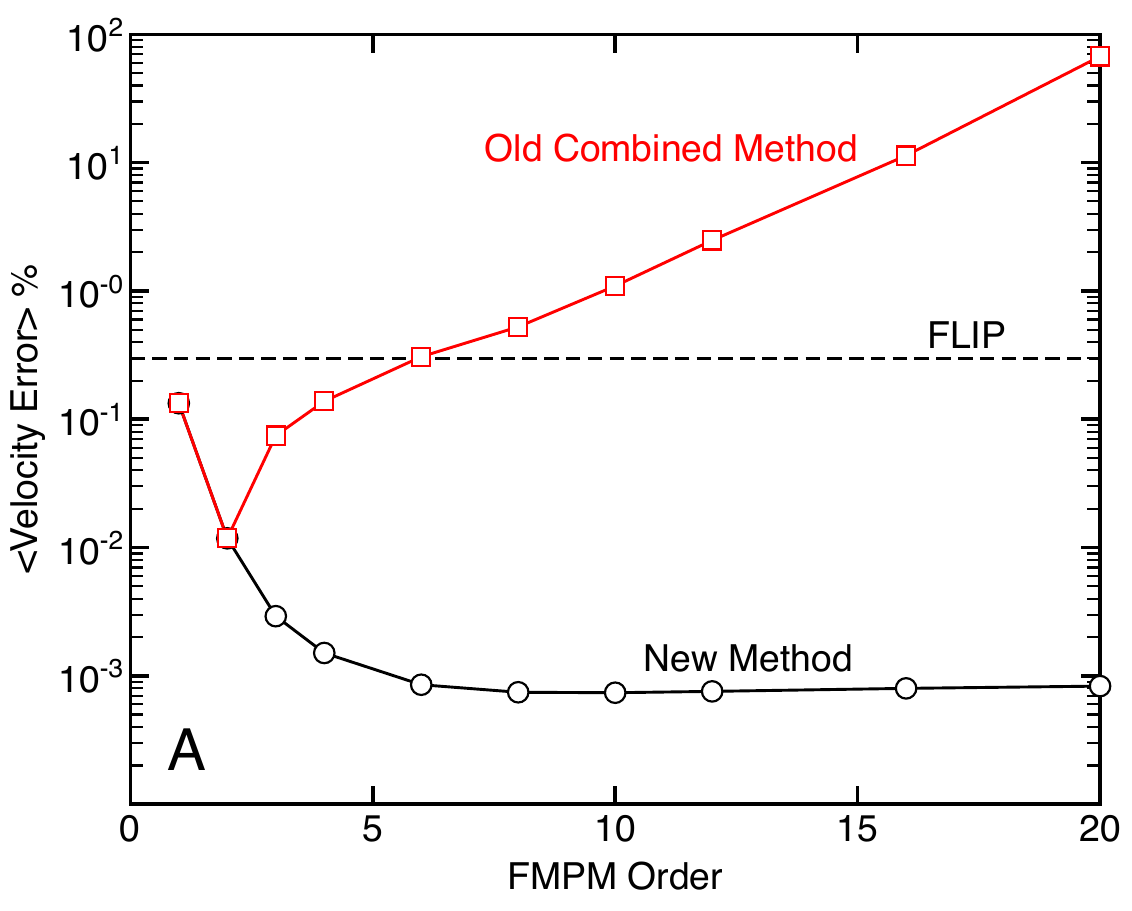}
\hspace{0.02\linewidth}
\includegraphics[width=0.45\linewidth]{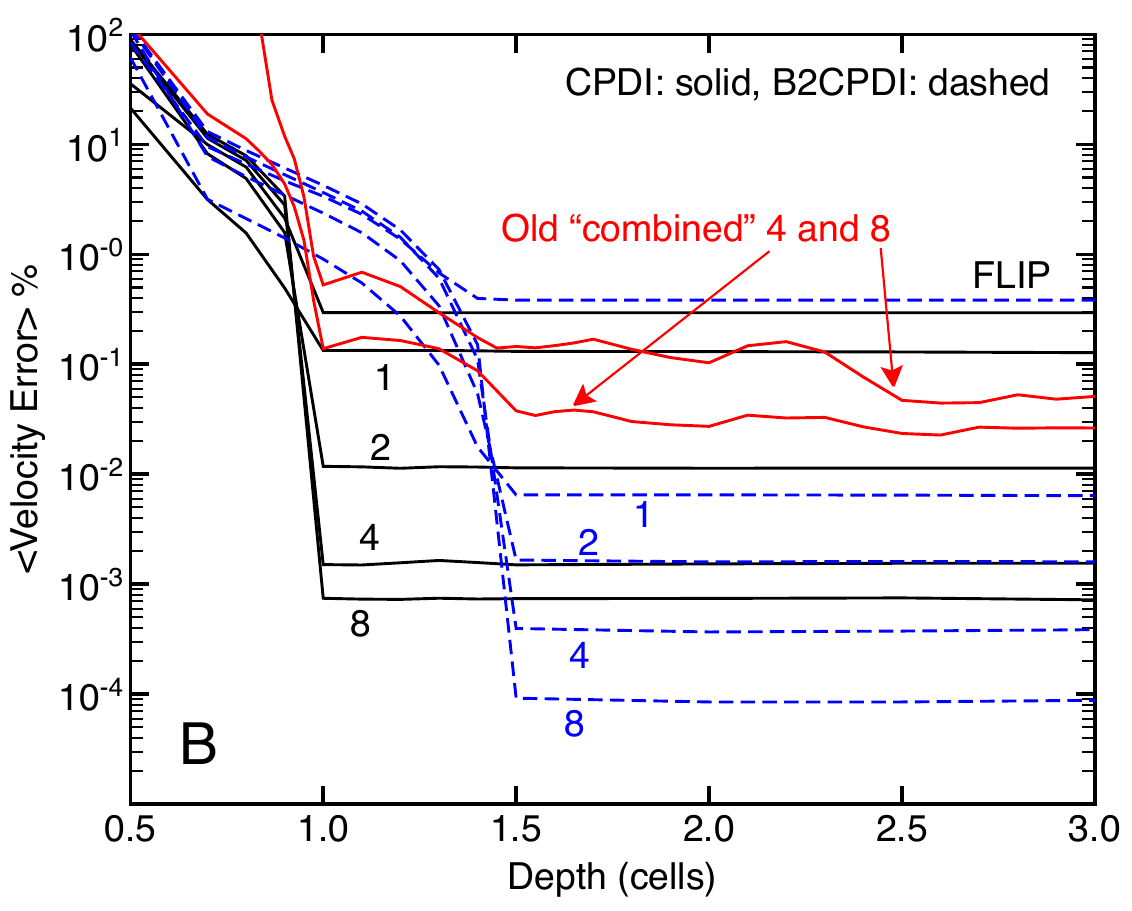}
\caption{A. Error of FMPM($k$) calculations as function of order $k$ using the new methods for velocity BCs and the old combined method from Ref.~\cite{Nairn:198}. B. Error in calculations as a function of depth of the wall in the moving-wall velocity BCs for FLIP and FMPM($k$) for $k=1$, 2, 4, and 8. Solid lines used CPDI shape function and dashed lines using B2CPDI shape functions. The solid, red lines are the old combined method for $k=4$ and 8 (note that $k=1$ and 2 are the same for new method and old combined method)}
\label{BCTest}
\end{figure}

Figure~\ref{BCTest}A compares results using CPDI shape function and wall depth of $d=1$ cell (0.25~m) for the new method, the prior ``combined'' method, and standard FLIP calculations. As expected, the new method and combined method were identical for $k\le2$, but diverged for $k>2$. The ``combined'' method continued to degrade with errors exceeding FLIP for $k>6$. In contrast, the new method maintained low errors for all $k$. Although it stopped improving for $k>8$, the average RMS error in that limit was 400 times lower than FLIP analysis. 

Figure~\ref{BCTest}B reprises a calculation from Ref.~\cite{Nairn:198} to evaluate moving wall conditions as a function of depth $d$. The new method (solid lines) results had errors that were constant for $d\ge1$, were much lower than FLIP, and kept decreasing as $k$ increased. The old ``combined'' method (solid black line for $k=2$ and wavy red lines for $k=4$ or 8) had errors that varied with $d$, were comparable to FLIP, and now \emph{increased} as $k$ increased. The dashed lines show the new method using B2CPDI shape functions (\emph{i.e.}, CPDI method calculated using quadratic spline grid functions). While FLIP errors were similar for CPDI and B2CPDI, FMPM($k$) showed even lower errors for B2CPDI. The errors for FMPM(8) using B2CPDI were 4400 times lower than FLIP errors. Importantly, however, good grid velocity conditions when using B2CPDI required $d\ge1.5$ cells to account for quadratic spline shape function support extending 1.5 cells from particle locations \cite{Steffen:2008,Nairn:198}.

\vskip6pt
\noindent \emph{Remark}: The above discussion commented that solving for effect of velocity BCs coupled through a full mass matrix is not possible. This new method does not find that solution, but it does allow FMPM($k$) to start with a lumped-mass velocity BCs and use additional calculations to keep the velocity BCs satisfied. The results were demonstrated to be much more accurate than FLIP calculations with the same BCs. This new algorithm removes all limitations of using FMPM($k$) in simulations with velocity BCs.

\subsection{Multimaterial Contact}\label{MMContact}

Standard calculations for MPM contact in steps 1.a and 3.a in \Fig{MPMFlowChart} find $\Delta\vec p^{(1)}$ and $\Delta\vec p^{(2)}$ using lumped mass methods. Like for velocity BCs, combining lumped mass contact methods with FMPM($k$) can cause conflicts. This issue was first noticed as artifacts in shock waves passing through material interfaces \cite{Nairn:198,Nairn:190}. An approximate method to fix this artifact was proposed for either XPIC($k$) \cite{Nairn:190} or FMPM($k$) \cite{Nairn:198}. That method, however, was for prior FMPM($k$) methods. This section derives contact methods that work for revised FMPM($k$) methods and also improves on the previous approximate method.
 
When using multimaterial MPM methods and FMPM($k$), each material extrapolates to its own velocity field and has its own approximate full mass matrix inverse. We assume two materials $\alpha$ and $\beta$. The initial extrapolations will find  separate momenta, $\vec p^\alpha$ and $\vec p^\beta$, full mass matrices, $\mfull^\alpha$ and $\mfull^\beta$, and lumped mass matrices, $\tens m^\alpha$, and $\tens m^\beta$. The terms are related to total momentum, full mass matrix, and lumped mass matrix by:
\begin{equation}
         \vec p = \vec p^\alpha + \vec p^\beta,\qquad \mfull = \mfull^\alpha + \mfull^\beta, \qquad {\rm and} \qquad \tens m = \tens m^\alpha+\tens m^\beta
\end{equation}
When using FMPM($k$), the FMPM loop is done for each material velocity field. In other words, grid velocities for the materials are:
\[
    \vec v^\alpha = (\mfull^\alpha)^{-1}\vec p^\alpha \quad {\rm and} \quad \vec v^\beta = (\mfull^\beta)^{-1}\vec p^\beta
\]

When these two materials move into contact, these velocities have to be adjusted, ideally to accurately reflect some model for contact mechanics. The goal is to change their velocities to $\vec v^\alpha + \Delta \vec v^\alpha$ and $\vec v^\beta+\Delta\vec v^\beta$, where relative motion is restricted on each node $i$ to be tangential to the contacting surfaces:
\[
\vec v^\beta_i+\Delta\vec v^\beta_i - (\vec v^\alpha_i + \Delta \vec v^\alpha_i) = k_i\hat{\vec t}_i \quad {\rm or} \quad
         \Delta\vec v_i^\beta - \Delta\vec v_i^\alpha  = k_i\hat{\vec t}_i  - \left(\vec v^\beta_i - \vec v^\alpha_i \right) 
\]
where $\hat{\vec t}_i$ is a unit vector tangential to contacting surface in the direction of motion on node $i$ and $k_i$ depends on contact law. To conserve momentum, these velocity changes should be related to equal and opposite momenta added to material momenta such that $\vec p^\alpha \to \vec p^\alpha+\Delta\vec p$ and $\vec p^\beta \to \vec p^\beta-\Delta\vec p$ resulting in
\begin{equation}
         \Delta\vec v^\alpha_i = \bigl((\mfull^\alpha)^{-1}\Delta \vec p\bigr)_i \qquad {\rm and} \qquad  \Delta\vec v^\beta_i = -\bigl((\mfull^\beta)^{-1}\Delta \vec p\bigr)_i
\end{equation}
By defining a reduced full mass matrix as $\mredfull = \left((\mfull^\alpha)^{-1} + (\mfull^\beta)^{-1}\right)^{-1}$,
the contact-induced momentum change for nodes in contact becomes
\begin{equation}
     - \bigl((\mredfull)^{-1}\Delta\vec p\bigr)_i = k_i\hat{\vec t}_i - \left((\mfull^\beta)^{-1}\vec p^\beta - (\mfull^\alpha)^{-1}\vec p^\alpha \right)_i
     \label{fmmcontact}
\end{equation}
Like the comparable equation for velocity BCs (see \Eq{CoupledBCs}), this equation provides $n_c$ equations (where $n_c$ is number of contact nodes) to find the $n$ unknowns in $\Delta \vec p$.

Because full-mass-matrix contact calculations appears infeasible (\emph{i.e.}, under determined), the goal changes to finding the best option for using lumped-mass contact calculations in simulations using FMPM($k$). The first step is to express lumped-mass contact calculations in general terms. For any node $i$ in multimaterial MPM that has more than one material velocity, contact calculations start by finding contact normal, $\hat{\vec n}_i$, and normal direction separation distance, $\delta_n$. The recommended method is logistic regression because it finds an accurate $\hat{\vec n}_i$ and can account for particle deformation when finding $\delta_n$ \cite{Nairn:190}. Next, find the momentum change needed for the two materials to move to the lumped-mass, center-of-mass velocity field as:
\[
         \Delta \vec p_i(0) = \frac{m_i^\alpha\vec p_i^\beta - m_i^\beta\vec p_i^\alpha}{m_i^\alpha +m_i^\beta} 
\]
The implied pressure at the contact surface is then $N_c = -\Delta \vec p_i(0)\cdot \hat{\vec n}_i/(A_c\Delta t)$ where $A_c$ is contact area and $\Delta t$ is the time step. A node is then determined to be contact if and only if both $N_c>0$ and $\delta_n<0$.\footnote[2]{Early MPM contact methods based only on $N_c>0$ or attempts to use only $\delta_n<0$ give inferior results to methods that combine the two contact criteria.} The first is required to find the contact surfaces in compression while second is required to find the surfaces actually touching. By lumped-mass matrix special case of \Eq{fmmcontact}, the momentum change to restrict surfaces to tangential motion is;
\[
    \Delta \vec p_i  =  \Delta \vec p_i(0)-m^{(red)}_i  k_i\hat{\vec t}_i \quad {\rm and} \quad m^{(red)}_i=\frac{m_i^\alpha m_i^\beta}{m_i^\alpha+m_i^\beta}
\]
Implementing contact mechanics follows by interpreting $\Delta \vec p_i$ as applying a tangential force:
\[
         \vec f_i^{(c)} = \frac{  \Delta \vec p_i\cdot \hat{\vec t}_i}{\Delta t} =  S_{slide}A_c = \frac{ \Delta \vec p_i(0)\cdot \hat{\vec t}_i - m^{(red)}_i k_i }{\Delta t}
\]
where $S_{slide}$ is tangential traction and $A_c$ is the contact area. This approach can implement numerous contact laws by defining how $S_{slide}$ depends on $N_c$ and potentially on any other parameters such as sliding velocity.  Solving for $k_i$ completes the analysis with:
\begin{equation}
       k_i = \frac{\Delta \vec p_i(0)\cdot \hat{\vec t}_i - S_{slide}A_c\Delta t}{m^{(red)}_i} 
       \quad \implies \quad  \Delta \vec p_i = (\Delta \vec p_i(0)\cdot \hat{\vec n}_i)\hat{\vec n}_i + S_{slide}A_c\Delta t\thinspace\hat{\vec t}_i
       \label{lumpedDelPi}
\end{equation}
The first term applies normal component of $\Delta \vec p_i(0)$ to remove interpenetration while the second is tangential force, which is zero for frictionless or non-zero for other contact laws. Nodes not in contact set $\Delta\vec p_i=0$.

The goals for contact methods in FMPM($k$) are to a.) define a method that reverts to single material mode if $\Delta p_i$ is set to $\Delta p_i(0)$ for all multimaterial nodes, and b.) provide stable contact calculations. Three easy methods that were tried first were to do nothing (\emph{i.e.} do standard lumped mass contact calculation in steps 1.a and 3.a in  \Fig{MPMFlowChart}), do lumped mass matrix calculations after the FMPM($k$) calculations (\emph{i.e.} move step 3.a to after step 3.b in  \Fig{MPMFlowChart}), or to combine these two methods (\emph{i.e.} keep step 3.a and repeat it after step 3.b in  \Fig{MPMFlowChart}). Unfortunately, all three ``easy'' methods fail the first criterion --- they do not revert to single material model by applying $\Delta p_i=\Delta p_i(0)$.\footnote[3]{The combined method does revert to single material mode for FMPM(2) but not for any $k>2$.} These simple methods are therefore rejected as viable for FMPM($k$) contact calculations.

Motivated by success of implementing velocity BCs by new calculations on each velocity increment, this work sought an incremental approach to contact mechanics. The end result of incremental contact calculations in FMPM($k$) methods leads to material velocities:
\begin{equation}
      \vec v^\alpha(k) = \sum_{\ell=1}^k \left(\Delta\vec v^\alpha(\ell)^{(0)} + \frac{\Delta \vec p_i(\ell)}{m^\alpha}\right)
      \quad {\rm and }\quad
      \vec v^\beta(k) = \sum_{\ell=1}^k \left(\Delta\vec v^\beta(\ell)^{(0)} - \frac{\Delta \vec p_i(\ell)}{m^\beta}\right)
      \label{evolving}
\end{equation}
where $\Delta\vec v^\alpha(\ell)^{(0)}$ and $\Delta\vec v^\beta(\ell)^{(0)}$ are the velocity increments found in $\ell^{th}$ pass of the FMPM($k$) loop \emph{before} correcting for contact mechanics and $\Delta \vec p_i(\ell)$ is momentum changed needed to correct the $\ell^{th}$ increment for contact. Assuming contact corrections have been applied for increments 1 to $\ell-1$ (with 1 for initial lumped velocities and contact calculations), the velocities after increment $\ell>1$ can for expressed as :
\begin{align*}
     \vec v^\alpha(\ell) & = \vec v^\alpha(\ell-1) + \Delta\vec v^\alpha(\ell)^{(0)} + \frac{\Delta \vec p_i(\ell)}{m^\alpha}
              =  \vec v^\alpha(\ell-1)^{(0)} + \frac{\Delta \vec p_i^{(net)}}{m^\alpha} + \Delta\vec v^\alpha(\ell)^{(0)} + \frac{\Delta \vec p_i(\ell)}{m^\alpha} \\
      \vec v^\beta(\ell) & = \vec v^\beta(\ell-1) + \Delta\vec v^\beta(\ell)^{(0)} - \frac{\Delta \vec p_i(\ell)}{m^\beta}
               =  \vec v^\beta(\ell-1)^{(0)} - \frac{\Delta \vec p_i^{(net)}}{m^\beta} + \Delta\vec v^\beta(\ell)^{(0)} - \frac{\Delta \vec p_i(\ell)}{m^\beta} \\
\end{align*}
Here $\vec v^\alpha(\ell-1)$ (or $\beta$) are the evolving contact-corrected velocities or \Eq{evolving} with $k=\ell-1$, $\vec v^\alpha(\ell-1)^{(0)}$ (or $\beta$) are total uncorrected velocities, and $\Delta \vec p_i^{(net)}$ is sum the contact corrections for the increments 1 to $\ell-1$; namely
\[
        \vec v^\alpha(\ell-1)^{(0)} = \sum_{j=1}^{\ell-1}\Delta\vec v^\alpha(j)^{(0)} 
        \quad {\rm and} \quad
        \Delta \vec p_i^{(net)} = \sum_{j=1}^{\ell-1} \Delta \vec p_i(j)
\]

Two incremental methods were derived differing only by which velocities were input to contact calculations. The ``Evolving'' method restricts current (evolving) velocities to tangential motion or  $\vec v^\beta(\ell)-\vec v^\alpha(\ell)=k_i\hat{\vec t}_i$. The ``Net'' method instead restricts net uncorrected velocities to tangential motion or  $\vec v^\beta(\ell)^{(0)}-\vec v^\alpha(\ell)^{(0)}=k_i\hat{\vec t}_i$. By either method, the lumped mass analysis above can be reused with the only change being that $\Delta\vec p_i(0)$ changes too:
\begin{equation}
      \Delta\vec p_i(0) =  \Delta\vec p_i^{(prior)}(0) + \Delta\vec p_i^\ell(0)       \label{incDelPiZero}
\end{equation}
where $\Delta\vec p_i^{(prior)}(0)$ and  $\Delta\vec p_i^\ell(0)$ are $\Delta\vec p_i(0)$ to move to center-of-mass velocity based on previous velocities and the current uncorrected velocity increment, respectively:
\begin{align*}
      \Delta\vec p_i^{(prior)}(0) &= \left\{\begin{array}{ll}
                 m_i^{(red)}\bigl(\vec v^\beta(\ell-1)-\vec v^\alpha(\ell-1)\bigr) & {\rm Evolving} \\
                 m_i^{(red)}\bigl(\vec v^\beta(\ell-1)^{(0)}-\vec v^\alpha(\ell-1)^{(0)}\bigr) & {\rm Net}
                 \end{array}\right. \\
     \Delta\vec p_i^\ell(0) & = m_i^{(red)}\bigl(\Delta\vec v^\beta(\ell)^{(0)}-\Delta\vec v^\alpha(\ell)^{(0)}\bigr)
\end{align*}
In other words, $\Delta\vec p_i^{(prior)}(0)$ in ``Evolving'' or ``Net'' methods is momentum change needed to move previous evolving velocities or net uncorrected velocities, respectively, to center of mass velocities. Substituting $\Delta\vec p_i^{(prior)}(0)$ into \Eq{incDelPiZero} gives $\Delta\vec p_i(0)$ that is input to contact calculations. The $\Delta\vec p_i$ output of those calculations provides $\Delta\vec p_i(\ell)$ with $\ell$ from 1 to $k$. The algorithms for contact calculations by ``Evolving'' or ``Net'' methods are summarized as follows:

\begin{enumerate}

\item Before starting the FMPM($k$) loop implement standard lump-mass contact calculations by step 3.a in  \Fig{MPMFlowChart}

\begin{itemize}

\item \emph{Evolving}: To support incremental calculations, this method needs to track $\Delta\vec p_i^{(prior)}(0)$ on each contact node. It is initialized here to $\Delta\vec p_i^{(prior)}(0) =  \Delta\vec p_i(0)-\Delta\vec p_i$.

\item \emph{Net}: To support incremental calculations, this method needs to track both $\Delta\vec p_i^{(prior)}(0)$ and $\Delta\vec p_i^{(net)}$on each contact node. They are initialized here to $\Delta\vec p_i^{(prior)}(0) =  \Delta\vec p_i(0)$ and $\Delta \vec p_i^{(net)}=\Delta\vec p_i$.

\end{itemize}

For efficiency, this first step can also cache contact calculations that will not change during the FMPM($k$) loop (\emph{e.g.}, contact normal, contact surface separation, contact area, \emph{etc.}).

\item After finding $\Delta\vec v^\alpha(\ell)^{(0)}$ and $\Delta\vec v^\beta(\ell)^{(0)}$ for each FMPM($k$) increment, repeat contact calculations. For nodes in contact find $\Delta\vec p_i(\ell)$ as $\Delta\vec p_i$ in \Eq{lumpedDelPi} with $\Delta\vec p_i(0)$ from \Eq{incDelPiZero}. For nodes not in contact, set $\Delta\vec p_i(\ell)=0$. Continue for the two incremental methods with:

\begin{itemize}

\item \emph{Evolving}: Use $\Delta\vec p_i(\ell)$ to correct the current velocity increment and then update
\[ 
     \Delta\vec p_i^{(prior)}(0)=\Delta\vec p_i(0)-\Delta\vec p_i(\ell)
\]
(\emph{i.e.}, increment it by $\Delta\vec p_i^\ell(0)-\Delta\vec p_i(\ell)$). Note that nodes not in contact that return $\Delta\vec p_i(\ell)=0$ will not change incremental velocities but will set $\Delta\vec p_i^{(prior)}(0)=\Delta\vec p_i(0)$ (\emph{i.e.}, increment it by $\Delta\vec p_i^\ell(0)$).

\item \emph{Net}: Use $\Delta\vec p_i^{(e\mathit{ff})}(\ell) = \Delta\vec p_i(\ell)-\Delta \vec p_i^{(net)}$ to correct the current velocity increment (\emph{i.e.}, $\Delta\vec p_i(\ell)$ implements contact in total uncorrected velocity and therefore needs to be reduced by any prior momentum changes to correct incremental velocities). Next update $\Delta\vec p_i^{(prior)}(0) = \Delta\vec p_i(0)$ and $\Delta \vec p_i^{(net)}=\Delta\vec p_i(\ell)$. Note that nodes not in contact that find $\Delta\vec p_i(\ell)=0$ will change the current velocity increment whenever previous $\Delta \vec p_i^{(net)}\ne0$.

\end{itemize}

\end{enumerate}

The first task is to evaluate if these new incremental methods revert to single material mode when contact calculations on all multimaterial nodes return $\Delta\vec p_i=\Delta\vec p_i(0)$ (\emph{i.e.}, applies momentum change for all nodes to move to center-of-mass velocity field). By the ``Evolving'' method, the first step reverts initial velocities to lumped-mass single material values and sets $\Delta\vec p_i^{(prior)}(0)=0$. Each increment finds $\Delta\vec p_i(\ell)=\Delta\vec p_i(0)=\Delta\vec p_i^\ell(0)$, which reverts the incremental velocity to center-of-mass velocity and maintains $\Delta\vec p_i^{(prior)}(0)=0$. By the ``Net'' method, the first step reverts initial velocities to lumped-mass single material values and sets $\Delta\vec p_i^{(prior)}(0)=\Delta \vec p_i^{(net)}=\Delta\vec p_i(0)$. Each increment finds $\Delta\vec p_i(\ell)=\Delta\vec p_i(0) = \Delta\vec p_i^{(prior)}(0) + \Delta\vec p_i^\ell(0)$ and $\Delta\vec p_i^{(e\mathit{ff})} =  \Delta\vec p_i(\ell)-\Delta\vec p_i^{(prior)}(0) = \Delta\vec p_i^\ell(0)$, which reverts the incremental velocity to center-of-mass velocity and adjusts $\Delta\vec p_i^{(prior)}(0)$ and $\Delta \vec p_i^{(net)}$ to the new $\Delta\vec p_i(0)$. In summary, both incremental methods correctly revert to single material mode where contact applies $\Delta\vec p_i(0)$ to all multimaterial modes.

\subsubsection{Shock Wave Simulations}

Because prior shock wave simulations revealed contact issues for both XPIC($k$) and FMPM($k$), 
reversion to single material mode was confirmed by repeating those calculations. In brief, the shock wave simulation from Ref.~\cite{Nairn:198} modeled 50 × 2.5 mm nickel bar confined by walls on top, bottom, and impacted on the right (at $x=50$~mm) by a wall moving at
1840 m/sec (40\%\ of the modeled material's bulk wave speed). The MPM conditions were 2D, plane strain analysis,
$0.25\times0.25$~mm cells, four particles per cell, B2CPDI shape functions, USL update method, coupled heat conduction, and Courant number $C = 0.2$ (where Courant number or Courant–Friedrichs–Lewy factor \cite{Courant:1967:PDE:1662749.1662757}, which in MPM is fraction of cell crossed in single time step at the material's wave speed). The reader is referred to Ref.~\cite{Nairn:198} for more details. To test contact methods, the nickel bar was split at $x=40$~mm into two bars with identical properties and then modeled using multimaterial MPM with contact mechanics.

\begin{figure}
\centering
\includegraphics[width=0.6\linewidth]{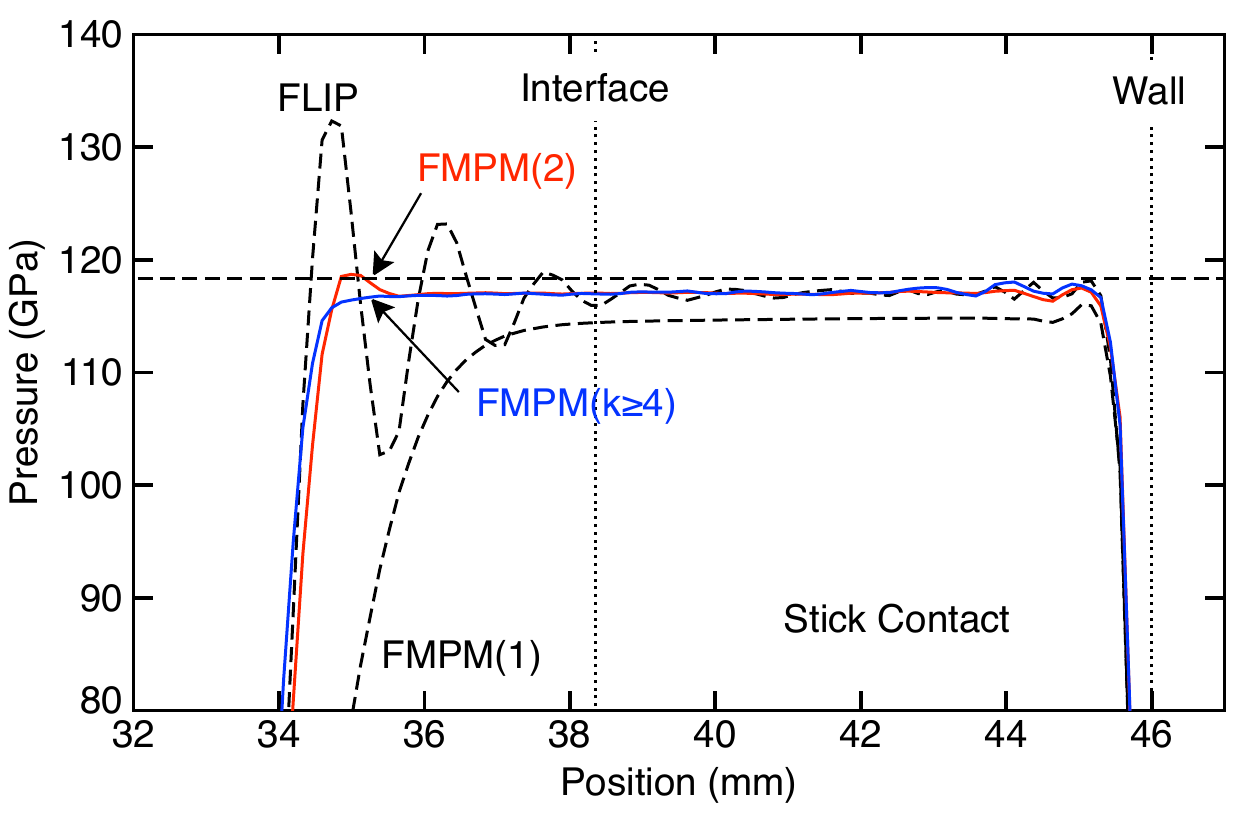}
\caption{Pressure shock wave passing a material interface using ``stick'' contact conditions by FLIP, FMPM(1), FMPM(2), and FMPM($k\ge4$); FMPM($k\ge2$) used either ``Evolving'' or ``Net'' methods with identical results. All results were also identical to simulations run in single material mode. The pressure wave is moving towards the left with leading edge a 34~mm and trailing edge near the wall at 46~mm. The dashed lines show the position of the material interface and the moving wall when the simulations was stopped. MPM results were extrapolated to the gird for plotting.}
\label{ShockStick}
\end{figure}

Figure \ref{ShockStick} plots calculations using ``stick'' contact (\emph{i.e.}, apply $\Delta\vec p_i = \Delta\vec p_i(0)$ to every multimaterial node which should revert to single material mode results). The results are for FLIP, FMPM(1), FMPM(2), and FMPM($k\ge4$) in multimaterial mode by either ``Evolving'' or ``Net'' methods and all are identical to calculations done in single material mode. Shock wave theory predicts a square wave with pressure 118.3~GPa. The FLIP results have large oscillations at leading edge of the pressure wave and that flattens out close to the theoretical value. Such oscillations are common in numerical methods trying to resolve square-wave edges. To reduce magnitude of such oscillations, all simulations used artificial viscosity \cite{Neumann:1950aa} with details in Ref.~\cite{Nairn:198}. The FMPM(1) results are over damped with the leading edge broadened by an unacceptable amount. FMPM(2) greatly reduces that broadening and gets good pressure wave with low oscillations, but still benefitted from artificial viscosity to keep oscillations low \cite{Nairn:198}. FMPM(4) differed from FMPM(2) only at the leading edge of the pressure wave. Finally, FMPM($k>4$) was within 0.5\%\ of FMPM(4) results for any value of $k$ confirming that these incremental contact calculations revert to single material mode for any FMPM($k$) order.

\begin{figure}
\centering
\includegraphics[width=0.6\linewidth]{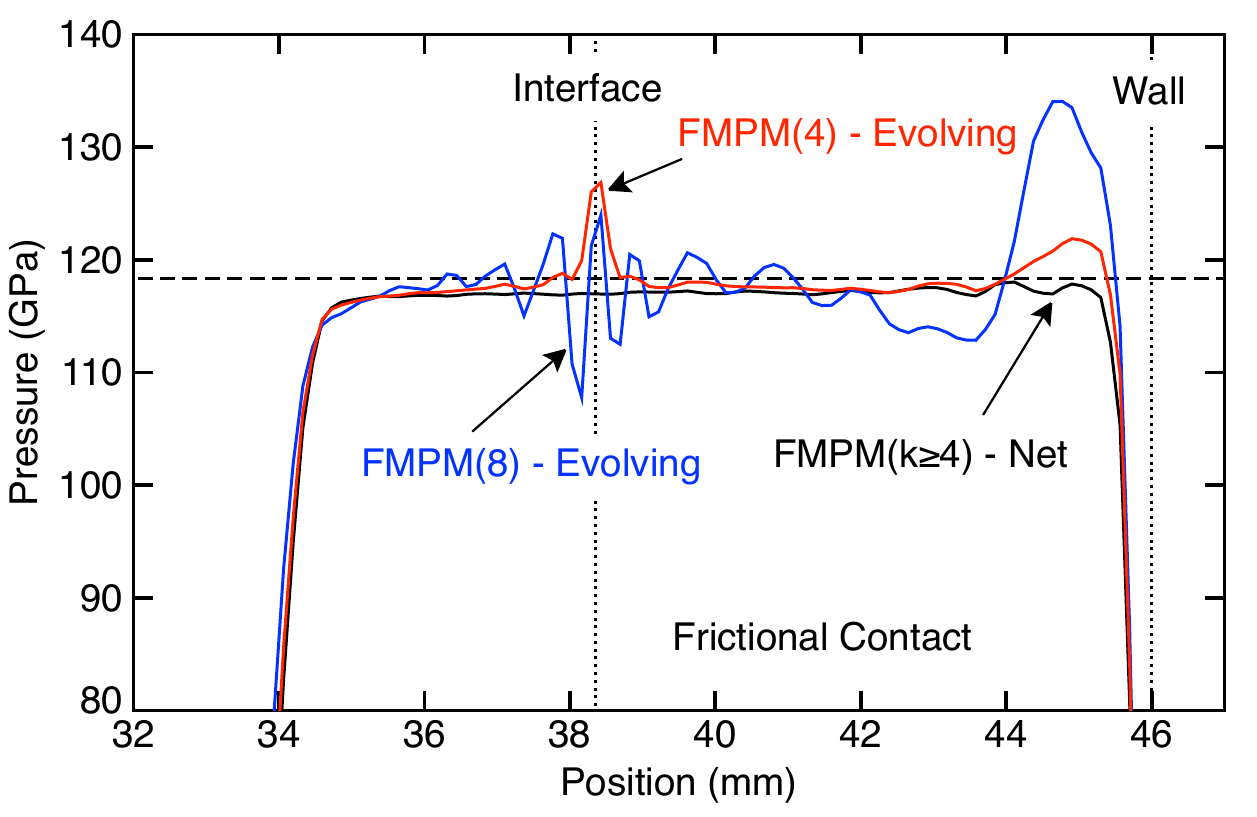}
\caption{Shock wave passing a material interface using frictional contact conditions by FMPM($k\ge4$) using the ``Net'' method and FMPM(4) or FMPM(8) using the ``Evolving'' method. The pressure wave is moving towards the left with leading edge a 34~mm and trailing edge near the wall at 46~mm. The dashed lines show the position of the material interface and the moving wall when the simulations was stopped. MPM results were extrapolated to the gird for plotting.}
\label{ShockFriction}
\end{figure}

Next, the shock simulations were repeated with a frictional interface. Because the interface is always under pressure, the results should again match single material mode. Any artifacts, such as wave reflections, at the interface, are therefore indications of inaccurate contact modeling. Because the top and bottom confining walls should eliminate tangential motion on the interface, the results are independent of the coefficient of friction.  Figure \ref{ShockFriction} gives results using FMPM($k\ge4$) by the ``Net'' method (and those results were within 0.5\%\ for all $k$) and FMPM(4) or FMPM(8) using the ``Evolving'' method. An FMPM(16) simulation by the ``Evolving'' method was unstable. The ``Net'' method works well while the ``Evolving'' method has unacceptable artifacts near the interface and at the wall that grow as $k$ increases.

To understand why the ``Net'' methods works, it suffices to consider FMPM(2) calculations and to assume a contact law where $S_{slide}A_c\Delta t = g(N_c)$ is a function only of compression on the interface. The ``Net'' method finishes the lumped-mass step with:
\[
        \Delta\vec p_i^{(prior)}(0) = \Delta\vec p_i(0) \quad {\rm and} \quad 
         \Delta\vec p_i^{(net)} = \Delta\vec p_i = (\Delta \vec p_i(0)\cdot \hat{\vec n}_i)\hat{\vec n}_i + g(N_c^{(1)})\hat{\vec t}_i
\]
where $N_c^{(1)}$ is contact pressure implied by $\Delta\vec p_i(0)$. For the FMPM increment in FMPM(2), the contact calculations result in:
\[
        \Delta\vec p_i^*(0) =  \Delta\vec p_i(0) + \Delta\vec p_i^{(2)}(0) \quad {\rm and} \quad
        \Delta\vec p_i(2) = (\Delta \vec p_i^*(0)\cdot \hat{\vec n}_i)\hat{\vec n}_i + g(N_c^{(2)})\hat{\vec t}_i
\]
where $N_c^{(2)}$ is contact pressure implied by $\Delta\vec p_i^*(0)$. The momentum change applied to the materials is
\[
         \Delta\vec p_i^{(e\mathit{ff})} = \Delta\vec p_i(2)-\Delta \vec p_i^{(net)}
         = (\Delta\vec p_i^{(2)}(0)\cdot \hat{\vec n}_i)\hat{\vec n}_i + g(N_c^{(2)})\hat{\vec t}_i - g(N_c^{(1)})\hat{\vec t}_i
\]
The total contact force becomes
\begin{equation}
       \vec f_i^{(c)} = \frac{(\Delta\vec p_i+\Delta\vec p_i^{(e\mathit{ff})})\cdot \hat{\vec t}_i}{\Delta t} = \frac{g(N_c^{(2)})\hat{\vec t}_i}{\Delta t} 
       \label{netfc}
\end{equation}
Because contact traction $S_{slide}A_c\Delta t$ is always based on total uncorrected velocity, which are reflected in each new $\Delta\vec p_i^*(0)$, this approach should work for any contact law --- including nonlinear laws or laws that depend on other parameters such as current velocity.

A comparable analysis for the ``Evolving'' method explains why it fails. The ``Evolving'' method finishes the lumped-mass step with:
\[
       \Delta\vec p_i = (\Delta \vec p_i(0)\cdot \hat{\vec n}_i)\hat{\vec n}_i + g(N_c^{(1)})\hat{\vec t}_i
            \quad {\rm and} \quad
       \Delta\vec p_i^{(prior)}(0) = \bigl(\Delta \vec p_i(0)\cdot \hat{\vec t}_i - g(N_c^{(1)})\bigr)\hat{\vec t}_i  = m_i^{(red)}k_i\hat{\vec t}_i
\]
For the FMPM increment in FMPM(2), the contact calculations reduce to:
\[
        \Delta\vec p_i^*(0) =  m_i^{(red)}k_i\hat{\vec t}_i+ \Delta\vec p_i^{(2)}(0) \quad {\rm and} \quad
        \Delta\vec p_i(2) = (\Delta\vec p_i^{(2)}(0)\cdot \hat{\vec n}_i)\hat{\vec n}_i + g(N_c^{(3)})\hat{\vec t}_i
\]
where $N_c^{(3)}=N_c^{(2)}-N_c^{(1)}$ is now contact pressure implied by $\Delta\vec p_i^{(2)}(0)$ or only by the incremental velocities. The total contact force becomes
\begin{equation}
       \vec f_i^{(c)} = \frac{\bigl(\Delta\vec p_i+\Delta\vec p_i(2)\bigr)\cdot \hat{\vec t}_i}{\Delta t} 
             = \frac{\bigl(g(N_c^{(1)}) + g(N_c^{(2)}-N_c^{(1)})\bigr)\hat{\vec t}_i}{\Delta t} 
              \label{evolvingfc}
\end{equation}
Comparing \Eq{netfc} for the ``Net'' method to \Eq{evolvingfc} for the ``Evolving'' method suggests they are equivalent provided $g(N_c)$ is linear (\emph{i.e.}, such that $g(N_c^{(2)}-N_c^{(1)}) =  g(N_c^{(2)})- g(N_c^{(1)})$).

\begin{figure}
\centering
\includegraphics[width=0.6\linewidth]{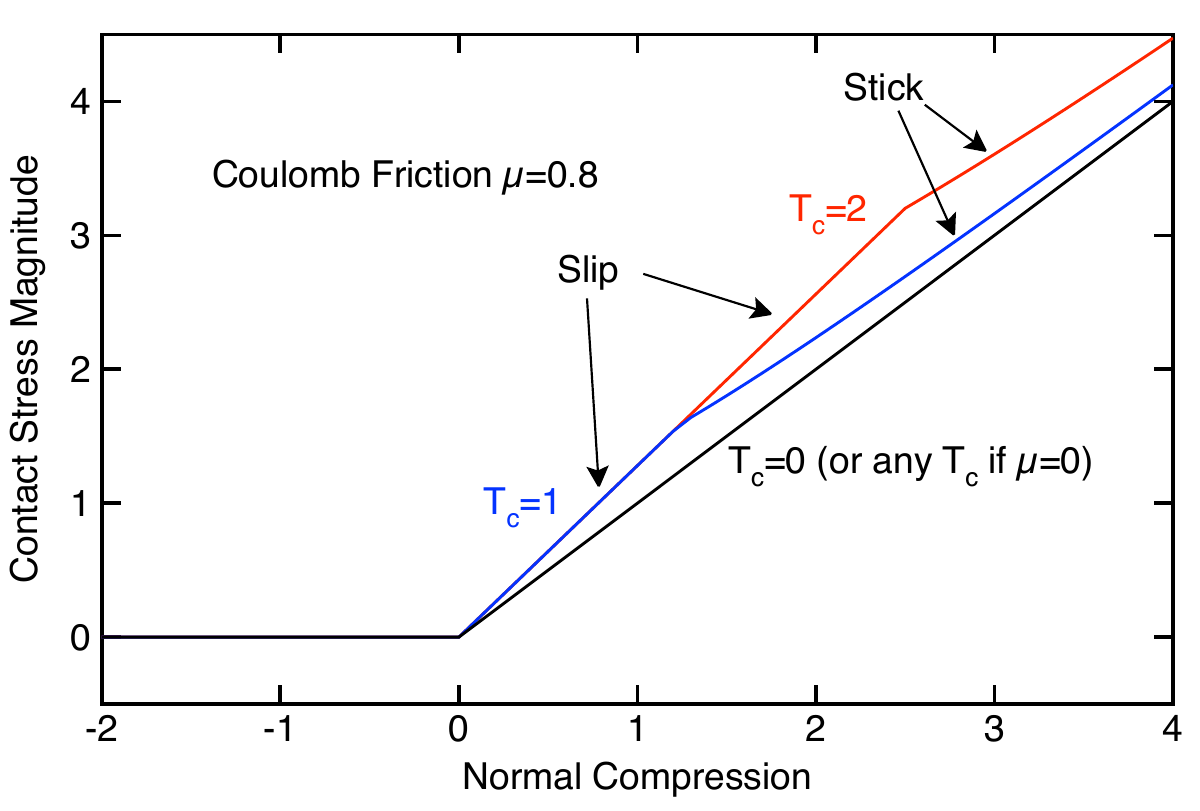}
\caption{A plot of $\|\vec f_c\|/A_c$ as a function of $N_c$ for three values of $T_c$ and $\mu=0.8$. Note that the plot for $T_c=0$ is also the plot for frictionless contact for all values of $T_c$.}
\label{CFLaw}
\end{figure}

Figure \ref{ShockFriction} used the frictionless contact law with $g(N_c)=0$, but the ``Net'' and ``Evolving'' methods still differed. The problem is that after including contact detection, all contact laws have nonlinear contact force $\vec f_i^{(c)}$. For example, modeling Coulomb friction finds frictional slip whenever the frictional traction, $\mu N_c$, is less than tangential traction needed to stick or $T_c =  \Delta \vec p_i(0)\cdot \hat{\vec t}_i/(A_c\Delta t)$ (note: direction of $\hat{\vec t}_i$ is selected to keep $T_c\ge0$). If the frictional traction exceeds tangential stick traction, then the interface sticks at the lower $T_c$ rather than slip with traction higher than needed to stick --- these two cases are described by $S_{slide}=\min(T_c,\mu N_c)$. Including contact detection, the magnitude of the contact force becomes
\[
        \frac{\|\vec f_c\|}{A_c} = \frac{\|\Delta\vec p_i\|}{A_c\Delta t} = \left\{\begin{array}{ll}
                   \sqrt{N_c^2 + \min(T_c,\mu N_c)^2} & N_c>0\ {\rm and}\ \delta_n<0 \\
                   0 & N_c\le0\ {\rm or}\ \delta_n\ge0
                   \end{array}\right.
\]
This contact model is plotted in \Fig{CFLaw} for three values of $T_c$ for contact node that has $\delta_n<0$. The non-contact region for $N_c\le0$ has zero contact force and it causes non-linearity in all contact laws.. For $T_c=1$ or $T_c=2$, the surfaces slip for low $N_c$ but then switch to stick when $\mu N_c$ exceeds $T_c$. For $T_c=0$, Coulomb friction with $\mu>0$ is linear for $N_c>0$, but still nonlinear when including all $N_c$. Note that the $T_c=0$ curve is also the contact force for any $T_c$ during frictionless contact with $\mu=0$ and all contact responses are slip. Thus, even frictionless contact has non-linear contact forces.

The ``Net'' method works because increments are based on $N_c^{(2)}$ from total uncorrected velocities meaning that contact is force is always sampling the same region of non-linear contact forces. In contrast, because $N_c^{(3)}=N_c^{(2)}-N_c^{(1)}$ is based only on incremental velocities and trends to smaller values, ``Evolving'' method increments can unrealistically move between regions of nonlinear plots such as switching from contact to non-contact or Coulomb friction switching from stick to slip. The ``Evolving'' method artifacts in \Fig{ShockFriction} are caused by increments finding loss of contact while a pressure wave should remain in contact. In contrast, $N_c^{(2)}$ remains high in the ``Net'' method and correctly maintains contact for all velocity increments. 

\subsubsection{Two Disks impact}

\begin{figure}
\centering
\includegraphics[width=0.8\linewidth]{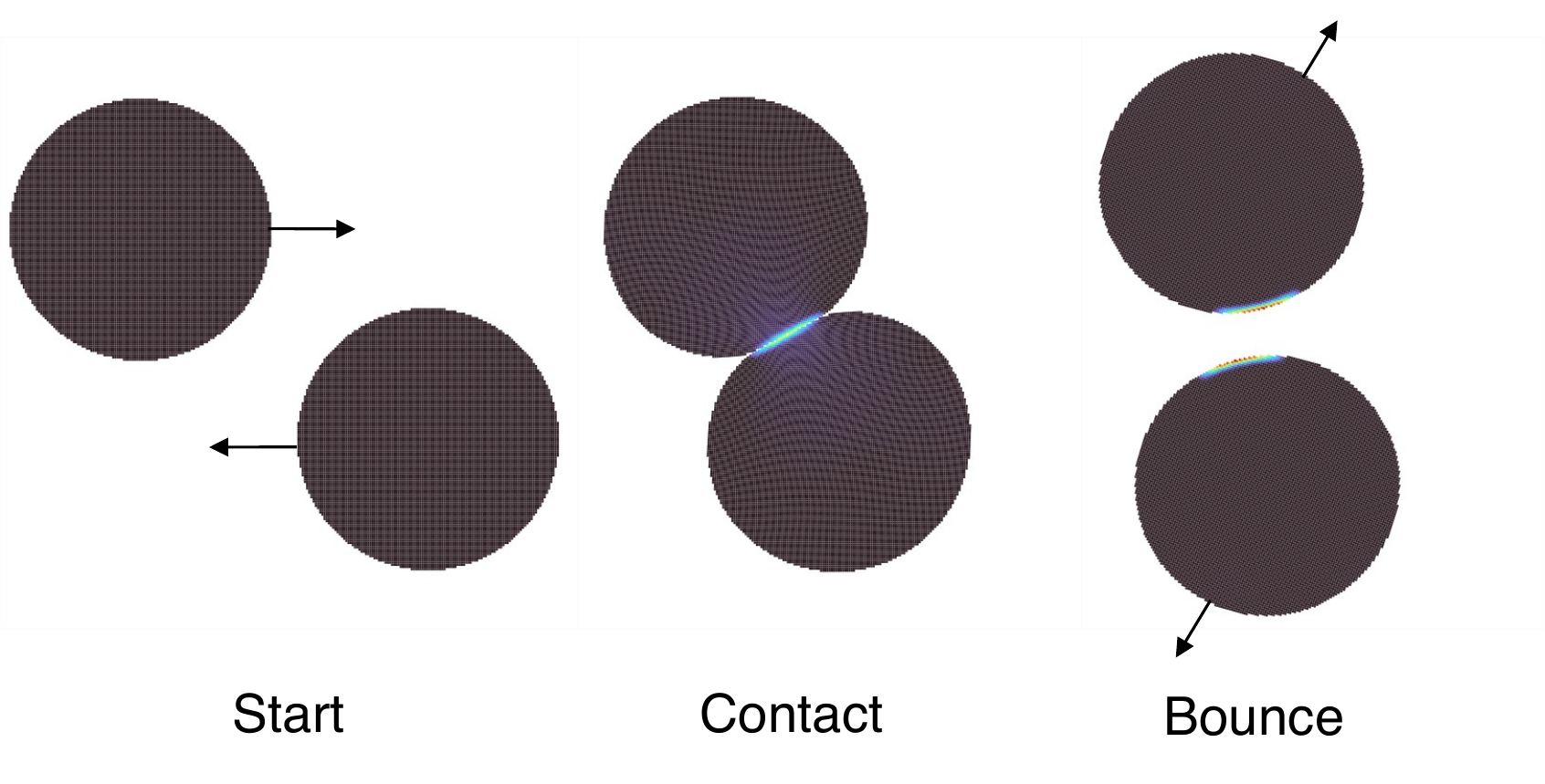}
\caption{Snapshots of two disks in off-center impact. The color is plotting temperature (from black for 300 K to red for 456 K). the cell size was 1/3~mm, coefficient of friction was $\mu=0.6$, and calculations used FMPM(4). Arrows indicate direction of motion.}
\label{TwoDisks}
\end{figure}

To test transitions from contact to non-contact as well slip-stick nonlinearities in Coulomb friction with coefficient of friction $\mu>0$, the next example modeling two disks (in 2D) moving toward each other with centers offset in the vertical direction (see \Fig{TwoDisks}). The offset impact causes the disks to change direction during impact and those changes are affected by $\mu$. Frictionless contact ($\mu=0$) will allow slippage and least amount of deviation. Increasing $\mu$ causes friction or stick forces that both change path of the objects and affects energy dissipation. The simulation details were two 20~mm diameter disks with centers separated 22~mm in $x$ direction and 16~mm in the $y$ direction moving toward each other with a velocity equal to 10\%\ of the material's wave speed or $v_x=81.65$~m/sec. Both disks were Neohookean materials (see \Eq{NHMat}) with $E=1000$~MPa and $\nu=0.33$ ($G=0.375.94$~Pa and $\lambda=729.77$) and $\rho=1.5$~g/cm$^3$. Contact calculations used logistic regression to find normal and separation. MPM details were USL, CPDI shape function, four particles per cell and ran for 0.16~ms such that impact finished and disks started moving apart.

\begin{figure}
\centering
\includegraphics[width=0.6\linewidth]{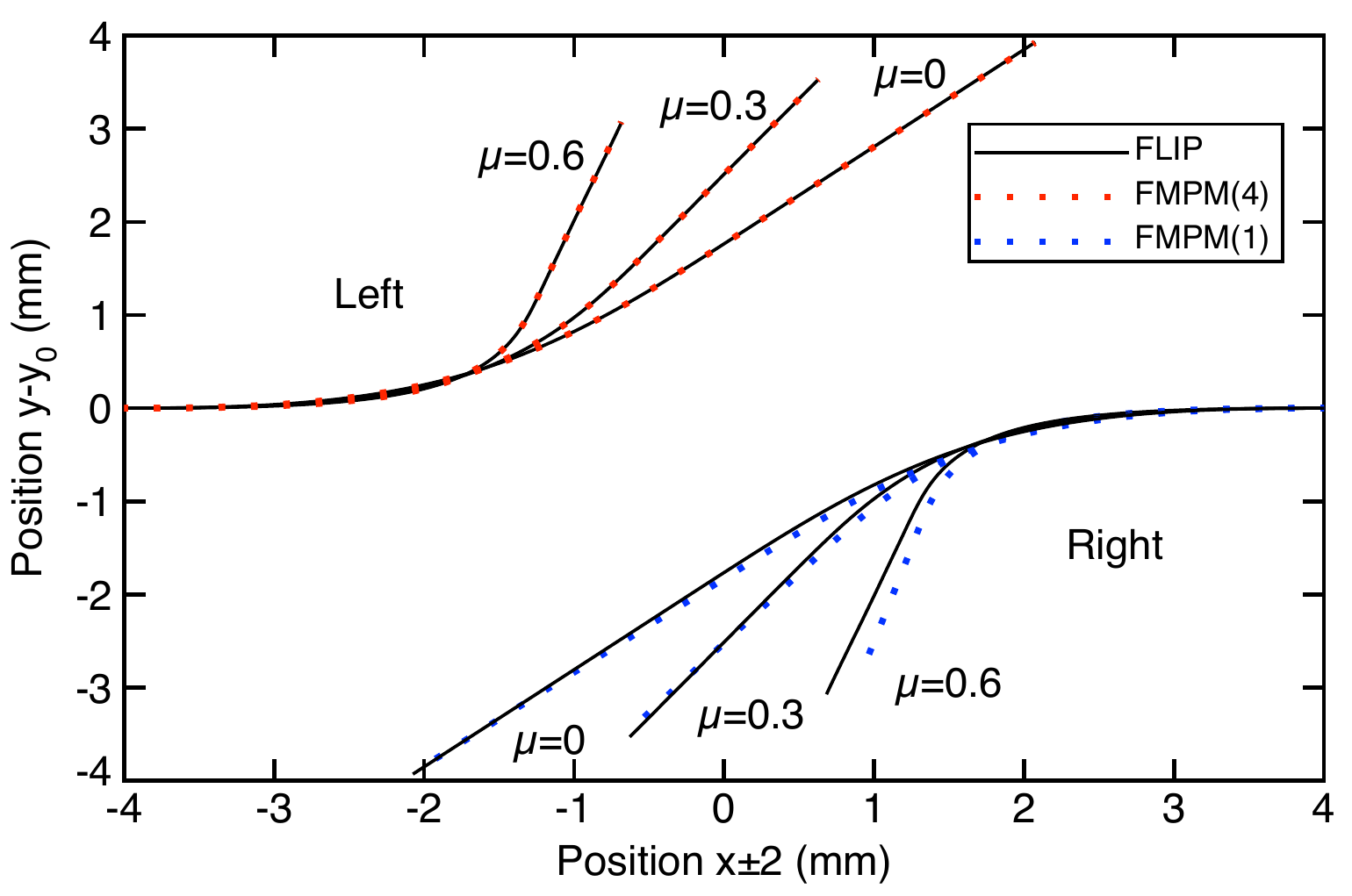}
\caption{The trajectories of the center of mass for left and right disks from 0 to 16~ms using FLIP (solid lines), FMPM(4) (red dots only for left disk) or FMPM(1) (blue dots only for right disk) for $\mu=0$, 0.3, and 0.6. Simulations used cell size = 1/3~mm (60 cells along diameter of the disks) and $C=0.2$. For clarity, the $y$ positions are relative to their starting positions and $x$ positions for left and right disks where shifted +2 or -2~mm, respectively.}
\label{DiskPaths}
\end{figure}

Figure \ref{DiskPaths} plots center-of-mass trajectory for each disk using FLIP, FMPM(1), and FMPM(4) for $\mu=0$, 0.3, or 0.6. As expected, the trajectories change direction during impact and the change becomes sharper as the $\mu$ increases. The black solid lines are for FLIP calculations. The FMPM(4) trajectories are superposed (with red dots) on the left disk results. The results were essentially identical to FLIP calculations. The FMPM(1) trajectories are superposed (with blue dots) on the right disk. These results showed slight deviation from FLIP results likely due to excess dissipation caused by FMPM(1). Results for FMPM($k>4$) matched the FMPM(4) results.

As expected, a sum of work energy ($W=\int_V \tens\sigma\cdot\tens\varepsilon$) and kinetic energy, $K$, reflects energy dissipation caused by friction work. A better approach to monitoring energy conservation is to use full-physics simulations that couples mechanical modeling to heat conduction with conversion of frictional work into heat energy. In addition, compressive stress in the Neohookean material can also be converted to heat. These simulations were therefore coupled to thermal transport modeling using material heat capacity $C_V=1500$~J/(kg-K) and thermal conductivity $\kappa=0.1$~W/(m-K) with stress-free temperature $T_0=300$~K. The colors in \Fig{TwoDisks} plot temperatures generated when $\mu=0.6$. The heating caused by friction and compression are highly localized on the contacting surfaces and did not conduct away much during the 16~ms simulation.

\begin{figure}
\centering
\includegraphics[width=0.45\linewidth]{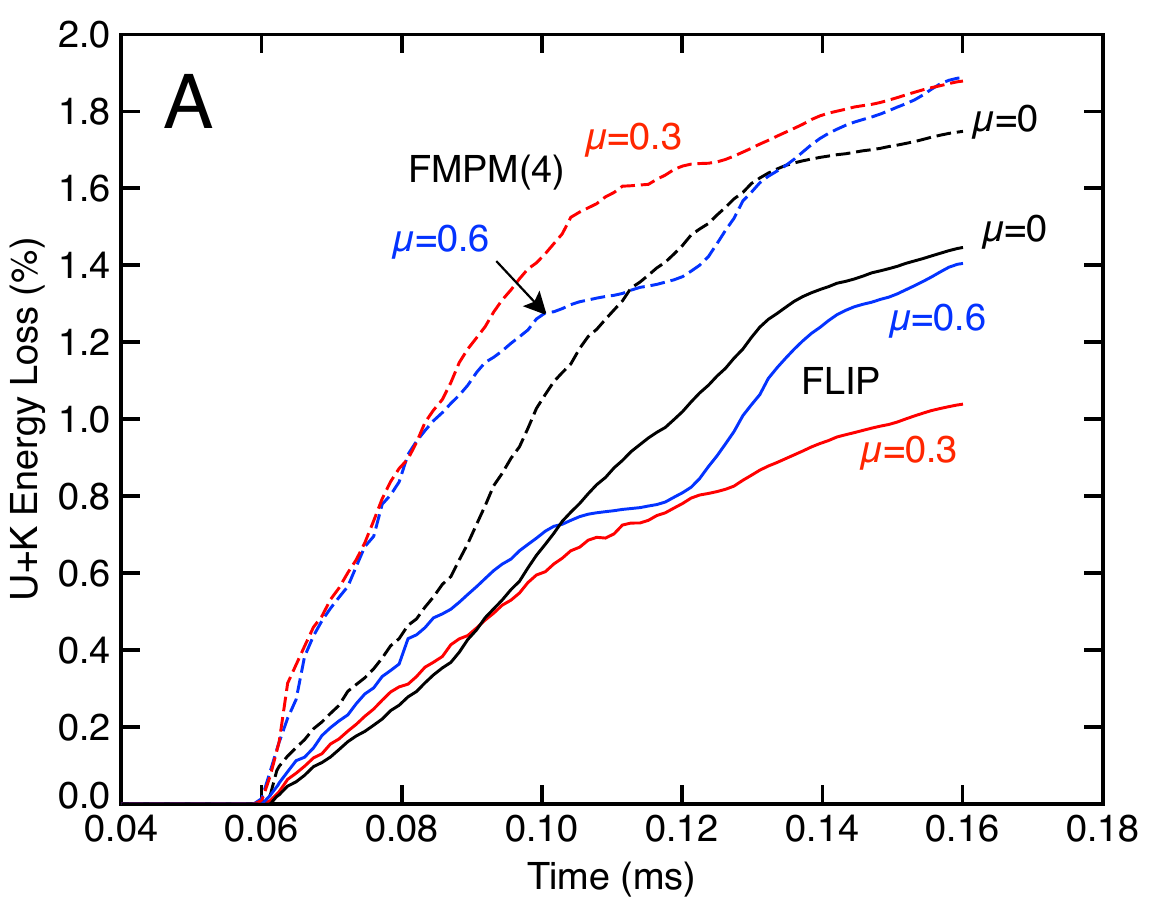}
\hspace{0.02\linewidth}
\includegraphics[width=0.45\linewidth]{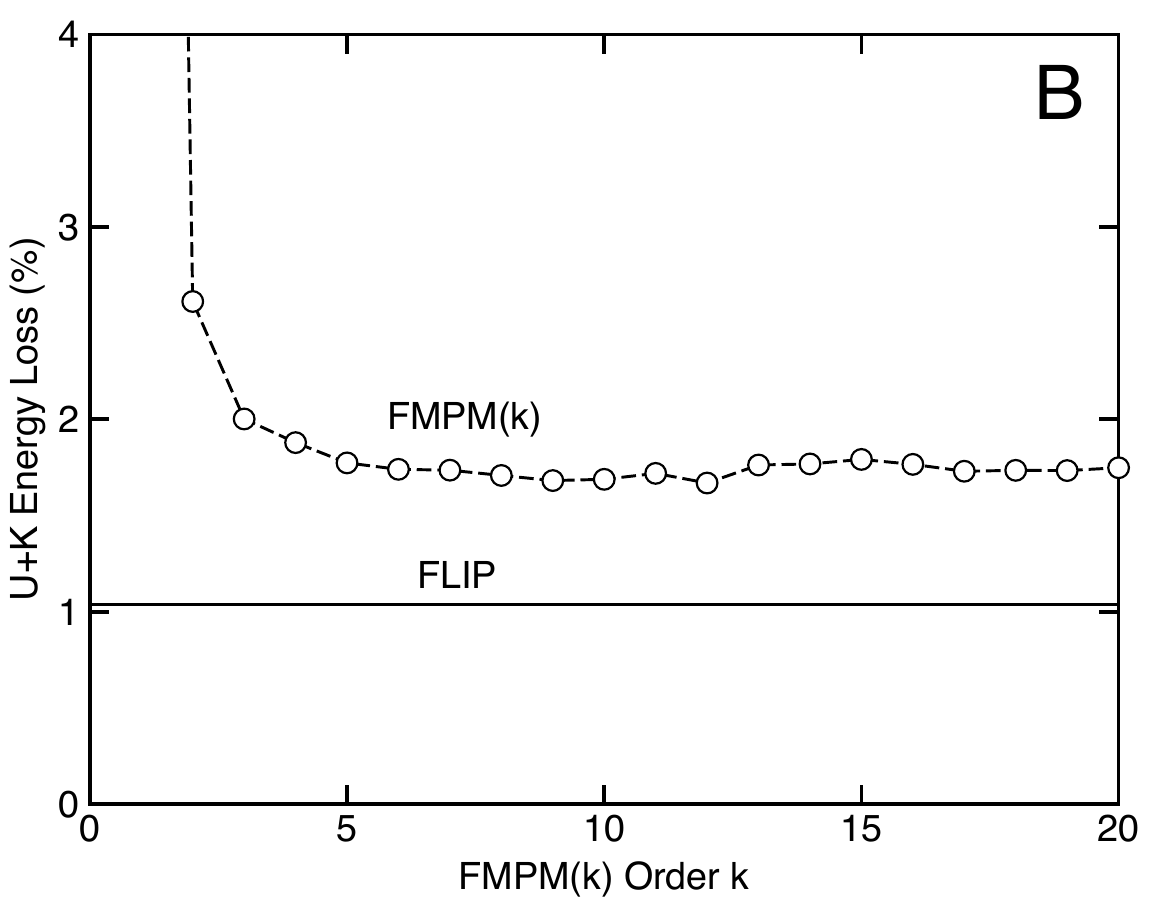}
\caption{Internal plus kinetic energy ($U+K$) energy loss at the end of the impact event (recorded at 16~ms) using cell size 1/3~mm and $C=0.2$. A. FLIP (solid lines) and FMPM(4) (dashed lines) results for $\mu=0$, 0.3, or 0.6. B. FLIP results (horizontal line) compared to FMPM($k$) as a function of $k$ using $\mu=0.3$.}
\label{UplusK}
\end{figure}

Because these simulations were globally adiabatic (no exchange of heat from the surroundings), the sum of internal energy, $U=W+Q$ ($Q$ is heat energy), and kinetic energy, $K$, should be conserved. Figure~\ref{UplusK}A plots $U+K$ energy lost by FLIP or FMPM(4) for $\mu=0$, 0.3, or 0.6. Both methods have small energy losses ($<2\%$) and losses were nearly independent of $\mu$. FMPM(4) losses were higher than FLIP, although still small. Figure~\ref{UplusK}B plots energy lost as a function of FMPM($k$) order $k$. The loss for FMPM(1) was very high (18.7\%). Using just FMPM(2) was a significant improvement. The energy loss continued to decrease but became nearly constant for $k>6$.

\begin{figure}
\centering
\includegraphics[width=0.6\linewidth]{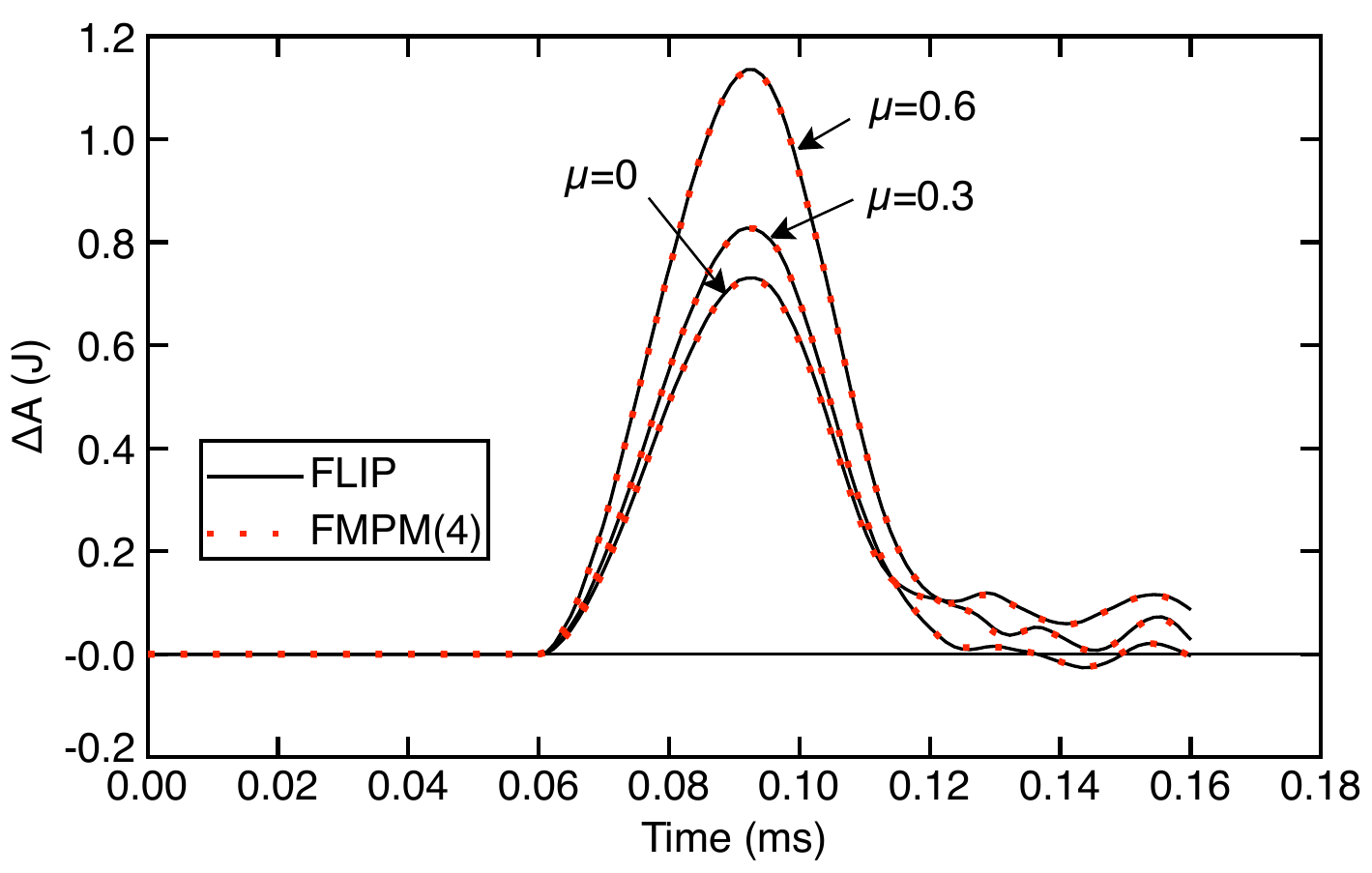}
\caption{Total Helmholz free energy as a function of time for cell size = 1/3~mm and $C=0.2$. The peak is caused by the impact event. The post-impact value returns close to zero with oscillations likely caused by remaining stress waves after the disks have fully separated (\emph{i.e.}, not an equilibrium state).}
\label{Helmz}
\end{figure}

A full-physics MPM simulation can also track entropy allowing calculation of Helmholz free energy, $A = U-TS$ (see details in Ref.~\cite{Nairn:2024}). An adiabatic process between two equilibrium states should find $\Delta A=0$. Figure~\ref{Helmz} plots $\Delta A = A(t)-A(0)$ as a function of time by FLIP or FMPM(4) for $\mu=0$, 0.3, or 0.6 . The FMPM(4) results (with red dots) are virtually identical to FLIP results. All results return close to $\Delta A=0$ after the impact but are not exactly zero and have some oscillations. These artifacts are likely due to residual stress waves in the disks after they have separated. The presence of oscillating stress waves means the final state has not yet returned to a new equilibrium state.

\begin{figure}
\centering
\includegraphics[width=0.45\linewidth]{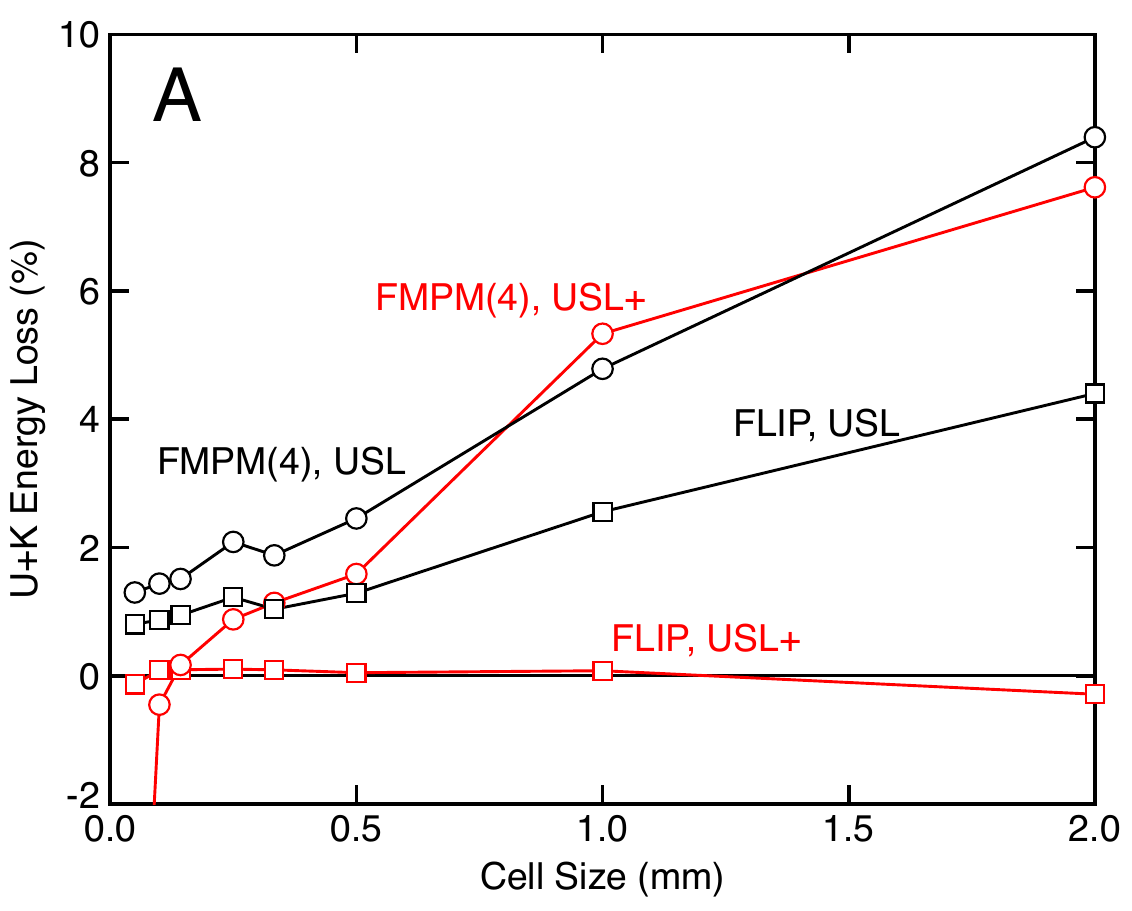}
\hspace{0.02\linewidth}
\includegraphics[width=0.45\linewidth]{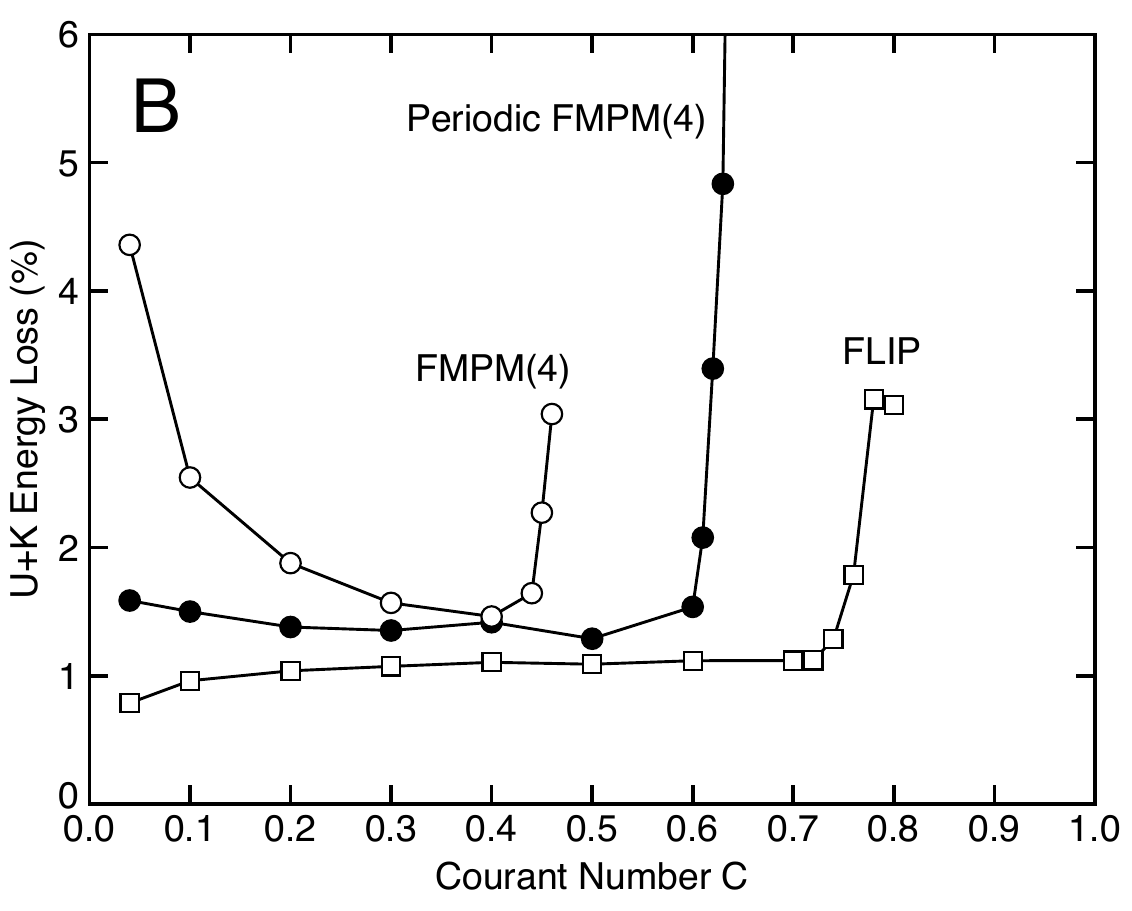}
\caption{Internal plus kinetic energy ($U+K$) energy loss at the end of the impact event (recorded at 16~ms) using $\mu=0.3$. A. Energy loss as function of cell size for $C = 0.2$ for FLIP or FMPM(4) with update methods USL or USL+. B. Energy loss as a function of Courant number for cell size 1/3~mm by FLIP, FMPM(4), or periodic FMPM(4).}
\label{UplusKConverge}
\end{figure}

The above results for disk impact all used cell size = 1/3~mm and $C = 0.2$. To test different settings, \Fig{UplusKConverge}A plots spatial convergence at constant Courant number $C$ while \Fig{UplusKConverge}B plots temporal convergence at constant cell size. Both used $\mu=0.3$. Because I have previously noticed improved energy conservation by using USL+, the spatial convergence results for FLIP and FMPM(4) used either USL or USL+. When using USL, both FLIP and FMPM(4) have typical convergence that approaches zero energy loss in the limit of zero cell size. As seen above in \Fig{UplusK}A, the FMPM(k) energy losses are higher than for FLIP. Their difference, however, decreases at smaller cell size. When using FLIP, the USL+ results are close to zero energy loss, but do not show monotonic convergence and developed negative loss at the highest resolution (non-physical energy increase). Because the differences between the USL and USL+ algorithms are small, it is surprising to find a large differences in convergence in these impact simulations. This lack for convergence with USL+ and an appearance of energy increases raises concern about the USL+ method. When using FMPM(4), the USL+ results are similar to USL for lower resolutions but degraded to energy increases at high resolution (reaching 7.8\%\ energy increase for cell size = 0.05 mm). Because USL+ is questionable when using FLIP and no better or worse when using FMPM(4), it should be avoided. Furthermore, USL+ should be avoided when using FMPM($k$) because it is less efficient (USL+ adds a second FMPM loop each time step compared to USL).

Figure \ref{UplusKConverge}B has temporal convergence for FLIP and FMPM($k$) as a function of Courant number $C$ revealing three key features:

\begin{enumerate}

\item Analysus of explicit methods often cites requirement that $C<1$ for stability. Explicit MPM approximately follows this rule but recent analysis suggests MPM needs $C\lesssim0.8$ \cite{Yang:2026aa}. The FLIP results in \Fig{UplusKConverge} agree with this reduced limit. They start degrade around $C=0.75$ and became unstable for $C>0.8$.

\item The FMPM(4) results show that it requires a lower $C<0.46$ (note: more analysis of FMPM($k$) stability is in the next section). FMPM(4) has a new problem that low $C$ causes energy losses to increase. The new issue is that using low $C$ at constant cell size means the simulation has more time steps. If the FMPM loop causes dissipation, even if only a small amount of dissipation, the addition of too many steps can add up and cause extra energy loss. Reference~\cite{Nairn:198} discusses this issue and proposes a solution whereby the FMPM($k$) calculations are done periodically rather then on every time step. Figure \ref{UplusKConverge}B includes results for periodic FMPM(4) where time step for FMPM($k$) calculations was based on constant $C_X=0.8$. Use of Periodic FMPM(4) solves the energy dissipation problem at low $C$\cite{Yang:2026aa} and extends stability limit to $C<0.6$.

\item Neither FLIP nor periodic FMPM(4) showed convergence as time step was made smaller. MPM simulations need to use $C$ lower enough to be stable but using much lower values provides little benefit.

\end{enumerate}

\vskip6pt
\noindent \emph{Remark}: The above contact methods do not provide improved contact modeling in MPM but rather provide improved FMPM($k$) results for problems with contact. Problems that are dominated by contact effects likely will find little benefit in using FMPM($k$) over FLIP. For example, the offset impact problem got good results with FLIP. Problems that have contact but can benefit for other reasons by using FMPM($k$) can use these revised methods to enable improved modeling. For example, FLIP can model a shock wave propagating through an interface, but develops large oscillations at the shock front (see \Fig{ShockStick}). Switching to FMPM($k)$ with revised contact continues to handle the interface and improves modeling of the shock front (see \Fig{ShockFriction}).

\subsection{FMPM($k$) Temporal Stability and Options to Improve Stability}

Numerical solutions to explicit differential equations require a sufficiently small time step for stable results. A rule-of-thumb is that the Courant number $C$ \cite{Courant:1967:PDE:1662749.1662757} must be less than 1. In explicit MPM modeling, $C$ is equivalent to fraction of a cell size that can be traversed in a single time step. Including supersonic simulations, where particle velocities might exceed stress wave speed, a choice for $C$ results in time step
\[
            \Delta t = C\frac{\Delta x_{min}}{\max\left(\vec v_{wave},\vec V_p\right)}
\]
where $\Delta x_{min}$ is the minimum cell size, $\vec v_{wave}$ is the material's wave speed, and $\vec V_p$ is the maximum velocity of any material point (simulations where $\vec v_{wave}$ or $\vec V_p$ vary need to dynamically adjust $\Delta t$ to retain stability). A recent  analysis of the energy behavior and spectral properties of single-step MPM updates concluded that a conservative estimate for maximum $C$ is between 0.71 and 0.77 \cite{Yang:2026aa}. This result matches the stability limit seen in disk-impact temporal convergences above in \Fig{UplusKConverge}. That figure also indicates that FMPM($k$) requires a smaller $C$ for stability. A possible reason is improved calculation of the mass-matrix inverse changes spectral properties of MPM updates. This section looks at stability of FMPM($k$) as a function or order $k$ and explores options for improving stability.

Yang and Sulsky \cite{Yang:2026aa} describe a freely-vibrating 1D bar that works well for testing stability limits. The left edge of the bar at $x=0$ is held fixed using zero-velocity boundary conditions. The bar extends to right edge at $x=L_r$ and undeformed particles are given initial velocity $v_x = v_0 \sin(\pi X_p/(2L_r))$. The bar should freely vibrate. Assuming a linear-elastic material, maximum bar extension occurs when all velocities drop to zero or all kinetic energy has been converted to strain energy. From that strain energy, the maximum extension corresponds to strain
\[
       \varepsilon_x = \frac{v_0}{v_{wave}}\cos\left(\frac{\pi x}{2L_r}\right)
\]
where $v_{wave}=\sqrt{E/\rho}$ is the material's wave speed. Integrating this ``resting'' strain, the maximum displacement at the bar end is
\[
        d_{max} = \frac{2L_rv_0}{\pi v_{wave}}
\]

A freely-vibrating bar simulation was used here to test FMPM($k$) stability in 1D MPM code with consistent units setting $L_r=40$, cell size 1 (with two particles per cell), material properties $E=2$ and $\rho=0.5$, and $v_0=0.16$. This velocity was chosen such that that $d_{max}=2$~cells (\emph{i.e.}, these stability calculations include stability of MPM encountering many cell crossings). These conditions resulted in vibrational period of 80 time units. Simulations ran for 5 periods or 400 time units. In general, the transition from stable to unstable was very sharp where a simulation completed 5 vibration periods with one $C$ but failed catastrophically when using $C$ that was 0.005 higher. These calculations thus varied $C$ and used manual binary searching to find $C$ stability limit within $\pm0.005$.

\begin{figure}
\centering
\includegraphics[width=0.45\linewidth]{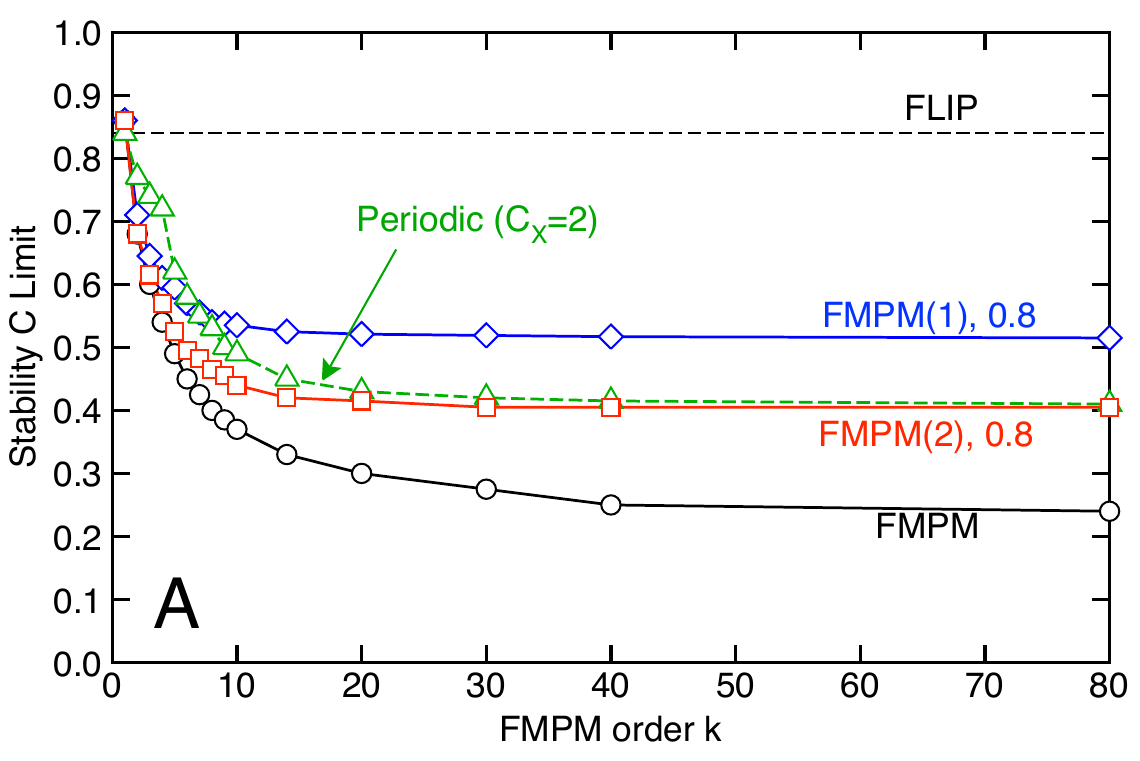}
\hspace{0.02\linewidth}
\includegraphics[width=0.45\linewidth]{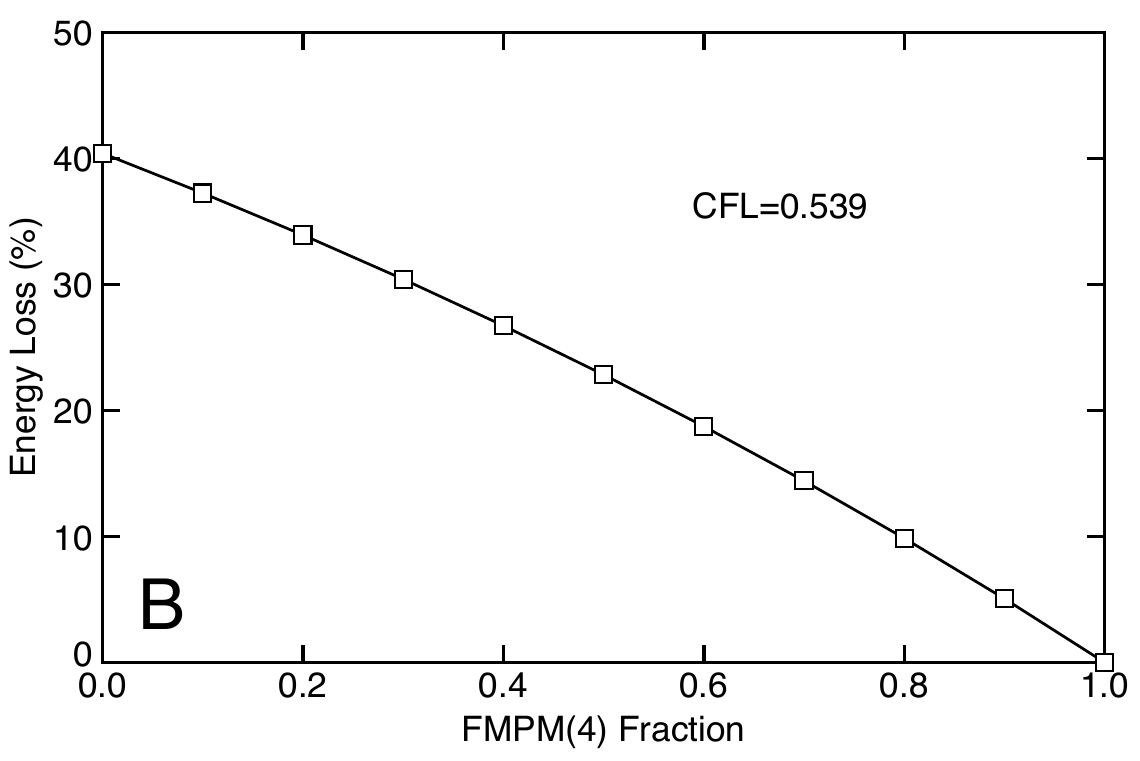}
\caption{A. Stability limit for FMPM($k$), FMPM($k$) blended with FMPM(1) using $\alpha=0.8$, FMPM($k$) blended with FMPM(2) using $\alpha=0.8$, and periodic FMPM($k$) using $C_X=2$ compared to stability limit for FLIP (dashed line at 0.84). B. Energy dissipation at the end of five vibrations using FMPM(4) as a function of fraction FMPM(4) (or $\alpha$) for fixed $C=0.539$.}
\label{FVStability}
\end{figure}

Figure \ref{FVStability}A plots $C$ limit for FMPM($k$) as a function of $k$ compared to FLIP limit, which is dashed line for $C=0.84$ (slightly higher than the conservative limit in Ref.~\cite{Yang:2026aa}). The FMPM($k$) stability limit drops rapidly with $k$. It does not, however, continue to drop. 1D calculations allowed FMPM($k$) up to $k=80$, which likely is never needed in MPM simulations. The stability limits for FMPM(40) and FMPM(80) were nearly identical (0.25 and 0.24). A conservative $C$ for any FMPM($k$) likely to be used in simulations is therefore $C=0.24$. Furthermore, the full mass matrix does not become ill conditioned, which should be reflected by a need for infinitesimally time steps or by maximum $C$ tending toward zero.

If an ill-conditioned full mass matrix does arise, an option to improve stability could be to blend the full mass matrix with a lumped mass matrix. First, replace the MPM mass matrix with a fraction $\alpha$ of the full mass matrix blended with a fraction $1-\alpha$ of the lumped mass matrix. The blended matrix become
\[
  \mfull^* = \alpha\mm\Sp\S + (1-\alpha) \mm = \mm\bigl(\tens I_n - \alpha(\tens I_n - \Sp\S)\bigr) = \mm\bigl(\tens I_n -\alpha\tens A\bigr)
\]
Comparing to \Eq{massfullform}, the only change to implement a blended mass matrix is to scale the $\tens A$ matrix by $\alpha$ leading to:
\[
       (\mfull^*)^{-1} = \bigl(\tens I_n + \alpha\tens A + \alpha^2\tens A^2 +\alpha^3\tens A^3 + \alpha^4\tens A^4 +\alpha^5\tens A^5 +\cdots\bigr)\mm^{-1}
\]
The only change needed to the FMPM loop in Table~\ref{FMPMcode} is to scale calculation of $\Delta\vec v(\ell)$ (in $\vec v_{prev}$) by $\alpha$. Figure \ref{FVStability}A gives $C$ limit plot for FMPM($k$) blended with FMPM(1) using $\alpha=0.8$. The limit for high $k$ is increased to $C=0.51$. The results using $\alpha=0.9$ gave limiting $C=0.40$ (not plotted).

Although blending with a lumped mass matrix enhances stability, it comes at the cost of excessive energy dissipation. All free vibration calculations for FLIP or non-blended FMPM($k$) for $k\ge2$ completed five vibration periods with the kinetic energy returning to within 0.03\%\ of the initial energy. FMPM(1), however, ended with 27.7\%\ dissipation using its stability limit of $C=0.86$. The problem with blending FMPM($k$) with FMPM(1) is that it blends a low-dissipation method with a high dissipation method. Figure \ref{FVStability}B plots energy dissipation of FMPM(4) as a function of blending fraction using $C=0.539$ (which is stability limit for non-blended FMPM(4)). The energy dissipation increases significantly as $\alpha$ decreases exceeding 40\%\ for $\alpha=0$  (this dissipation is higher than 27.7\%\ quoted above because that was for FMPM(1) using  $C=0.86$ while 40\%\ is the result when using $C=0.539$).

An option to gain stability by blending without causing excess dissipation is to blend the full mass matrix with the FMPM(2) mass matrix instead of the lumped mass matrix. The FMPM(2) matrix is derived by inverting its inverse:
\[
       \mfull_2^{-1} = \bigl(\tens I_n + \tens A\bigr)\mm^{-1} \quad \implies \quad \mfull_2 = \mm\bigl(\tens I_n +\tens A\bigr)^{-1}
\]
Blending this mass matrix with the full mass matrix gives
\[
      \mfull^* = \alpha\mm\bigl(\tens I_n - \tens A\bigr) + (1-\alpha)\mm \bigl(\tens I_n +\tens A\bigr)^{-1}
         = \mm\bigl(\tens I_n - \alpha\tens A^2\bigr)\bigl(\tens I_n +\tens A\bigr)^{-1}
\]
Expanding $(\mfull^*)^{-1}$ in a Taylor series gives
\[
       (\mfull^*)^{-1} = \bigl(\tens I_n + \tens A + \alpha\tens A^2 +\alpha\tens A^3 + \alpha^2\tens A^4 +\alpha^2\tens A^5 +\cdots\bigr)\mm^{-1}
\]
This new blending can be implemented in the FMPM loop by changing the recursion relation in \Eq{recursion} to
\[
       \Delta \vec v(\ell) = \alpha^{\ell\thinspace{\rm mod}\thinspace2}\tens A\Delta \vec v(\ell-1)
\]
In other words, scale the calculation of $\Delta\vec v(\ell)$ ($\vec v_{prev}$ in Table~\ref{FMPMcode}) by $\alpha$ only when $\ell$ is odd rather scale it for every $\ell$ as done to blend with a lumped mass matrix. Figure \ref{FVStability}A plots $C$ limit for FMPM($k$) blended with FMPM(2) for $\alpha=0.8$. The stability limit for high $k$ has increased to $C=0.40$ while keeping energy dissipation to within 0.03\%\ of the initial energy (and at no computational cost).

The approach to blending with FMPM(2) is easily generalized. To blend fraction $\alpha$ of FMPM($k$) with fraction $(1-\alpha)$ of FMPM($m$) simply change the recursion relation in \Eq{recursion} to 
\[
       \Delta \vec v(\ell) = \left\{ \begin{array}{ll}
              \alpha \tens A\Delta \vec v(\ell-1) & {\rm if}\ \ell\thinspace{\rm mod}\thinspace m = 1 {\rm\ or\ }m=1 \\
              \tens A\Delta \vec v(\ell-1) & {\rm otherwise}
              \end{array} \right.
\]
In other words, include scaling factor $\alpha$ every $m^{th}$ pass through the FMPM loop. Blending is characterized by two parameters --- $m$ and $\alpha$. Decreasing them enhances temporal stability but at a tradeoff of increased dissipation. Increasing them limits benefit of temporal stability. In the vibrating bar example, using $m=1$ caused too much dissipation for any non-zero $\alpha$ while using $m=2$ retained more than half of the benefit with no dissipation. The elimination of dissipation in the vibrating bar example was because FMPM(2) alone had no dissipation. If real-world problems, blending should likely use $m=2$ as significant reduction in dissipation compared to $m=1$ and than vary $\alpha$ in a tradeoff between enhanced stability and energy dissipation.

Another option to enhance stability is to use periodic FMPM($k$). For example, \Fig{UplusKConverge}A shows stability limit of periodic FMPM(4) in that impact example to be between normal FMPM(4) and FLIP. Figure \ref{FVStability}A plots $C$ limit from the vibrating bar example for periodic FMPM($k$) using $C_X=2$. The stability is improved without causing any dissipation. The stability limit for high $k$ is now $C=0.40$. Although periodic FMPM($k$) could use a higher $C$, the transition from stable to unstable became less sharp. The results in \Fig{UplusKConverge}A are based on maximum $C$ that had low dissipation after the five vibration periods. Some $C$ values between that value and the standard FMPM($k$) limit completed five vibration cycles but had increased noise in particle velocities.

Yet another potential option is to modify MPM to make use of velocity gradients in extrapolations \cite{Wallstedt:2007aa}. Preliminary results on extending velocity gradient methods for use in FMPM($k$) suggests that the stability limit for FLIP increases to $C=1$ and FMPM($k$) stability plateaus at $C=0.60$. Although few examples show compelling benefits of velocity gradient methods, a potential $2.5\times$ improvement in FMPM($k$) temporal stability might be reason enough to use them. This should be pursued in a future publication. 

A reader might wonder --- if FLIP is the most temporally stable, why ever use FMPM($k$)? The reason is that maximum $C$ is not the only factor in finding the best method or even the most stable method. One has to consider other factors relating to quality of simulation results. If maximum $C$ was the only factor, than FMPM(1) would be preferred because its maximum $C$ is 0.86 (not plotted in \Fig{FVStability}A). FMPM(1), however, is unacceptable because of excessive energy dissipation. A known problem with FLIP is development of null-space noise that can degrade simulations \cite{Nairn:179,ChrisNullSpace}. FMPM($k$) \cite{Nairn:198} and it predecessor XPIC \cite{Nairn:179}, were developed to reduce null-space noise. When such noise causes problems in a simulation, FMPM($k$) can provide enhanced overall stability, albeit with a need for more calculations and a smaller time step. The options of blending with FMPM(2), using periodic FMPM($k$), and potentially using velocity gradients all provide methods to enhance temporal stability of FMPM($k$).

\subsection{Dynamic FMPM}

The relative simulation time for FMPM($k$) is given in \Eq{compcost} \cite{Nairn:198}. Furthermore, the previous section shows that FMPM($k$) requires a smaller time step then FLIP. If FMPM($k$) gets good results in problems where FLIP fails, the computational cost is a secondary factor --- one should use the method that gives good results. Nevertheless, it is worth exploring options for improving the efficiency of FMPM($k$).

Revising \Eq{mfullEq}, the grid velocities found in the FMPM loop expand to
\begin{equation*}
      \vec v^+(k) = \vec v^+(1) + \Delta \vec v(2) + \Delta \vec v(3) + \Delta \vec v(4) + \cdots
\end{equation*}
where $\Delta \vec v(\ell) = \vec v^+(\ell) - \vec v^+(\ell-1)$ is the difference in finding velocity by FMPM($\ell$) and FMPM($\ell-1$). This form is a series of expansion for $\vec v^+(k)$ with terms $\Delta\vec v(\ell)$. An option to improve FMPM($k$) efficiency is to implement ``Dynamic FMPM'' whereby FMPM runs just until this series converges. The challenge is to implement a convergence criterion to decide when $\Delta\vec v(\ell)$ is sufficiently small.

This dynamic option was tested by reusing the problem of two 50×20 mm blocks impacting each other at 10\%\ of their wave speed in 2D, plane-strain simulations using CPDI shape functions, the USL update method, and $C= 0.2$ to keep stable for all orders tested (more details are in Ref.~\cite{Nairn:198}). The modeling used single-material mode (\emph{i.e.}, contact methods were not used) and had no other boundary conditions. The impact was used to induce vibrations and high velocity gradients and simulations can be evaluated by total energy dissipation after impact is complete (which should be zero for elastic materials). Figure 4 in Ref.~\cite{Nairn:198} shows the FMPM($k$) can get nearly a order of magnitude lower energy dissipation than FLIP and that dissipation continues to decrease as order increases. The relative simulation time, however, increased as order increases (fits to \Eq{compcost} for this problem using 8 processors and USL with $\phi=1$ found $F=0.244$). The question now is whether dynamic FMPM($k$) can  get similar improvements with shorter simulation times?

This simulation starts with all particles in each block moving at a constant velocity. Standard MPM exactly preserves such constant block motion (up until their velocity fields overlap). This fact means the round-trip extrapolations recover initial velocities, which translates to
\[
          \vec V = \tens S\tens S^{+} \vec V
\]
\emph{if and only if} all elements in $\vec V$ are constant (or at least constant within non-interacting blocks of material points). The first pass though an FMPM loop for a constant velocity block therefore finds
\[
         \Delta \vec v(2) = (\tens I - \tens S^{+}\tens S)\tens S^{+} \vec V = \tens S^{+} \vec V -  \tens S^{+}(\tens S\tens S^{+} \vec V) = 0
\]
As long as blocks are moving at constant velocity, the FMPM loop converges to correct answer in one pass.
Once the blocks interact, $\Delta\vec v(\ell)$ will not be zero and dynamic FMPM needs a convergence criterion. Two options tried for this example were
\[
         {\rm Means} = \frac{\sum_i ||\Delta\vec v_i(\ell)||}{\sum_i ||\vec v_i^+(\ell)||}, \quad {\rm and}\quad
         {\rm Changes} = \sum_i \frac{||\Delta\vec v_i(\ell)||}{||\vec v_i^+(\ell)|| + \epsilon}
\]
The ``Means'' criterion is ratio of mean magnitude of $\Delta\vec v_i(\ell)$ on the nodes to mean magnitude of full velocity $\vec v_i(\ell)$. The ``Changes'' is mean value for change in velocity on each node relative to velocity on that node (the $\epsilon$ term was used to avoid problems for nodes with zero velocity; it was set to 1\%\ of the maximum velocity magnitude).

Implementation of dynamic FMPM($k$) is straightforward. A simulation picks a maximum order $k$ and then inserts a convergence check each pass through the FMPM loop (see line ``(3)'' in Table~\ref{FMPMCode}). This line calls separate code to calculate a chosen convergence metric and compares to an input convergence criterion. If the metric is small enough, the FMPM loop exits; if not, the loop continues. To protect against non-convergence (or slow convergence), the loop exits if it reaches the assigned maximum order. 

\begin{figure}
\centering
\includegraphics[width=0.45\linewidth]{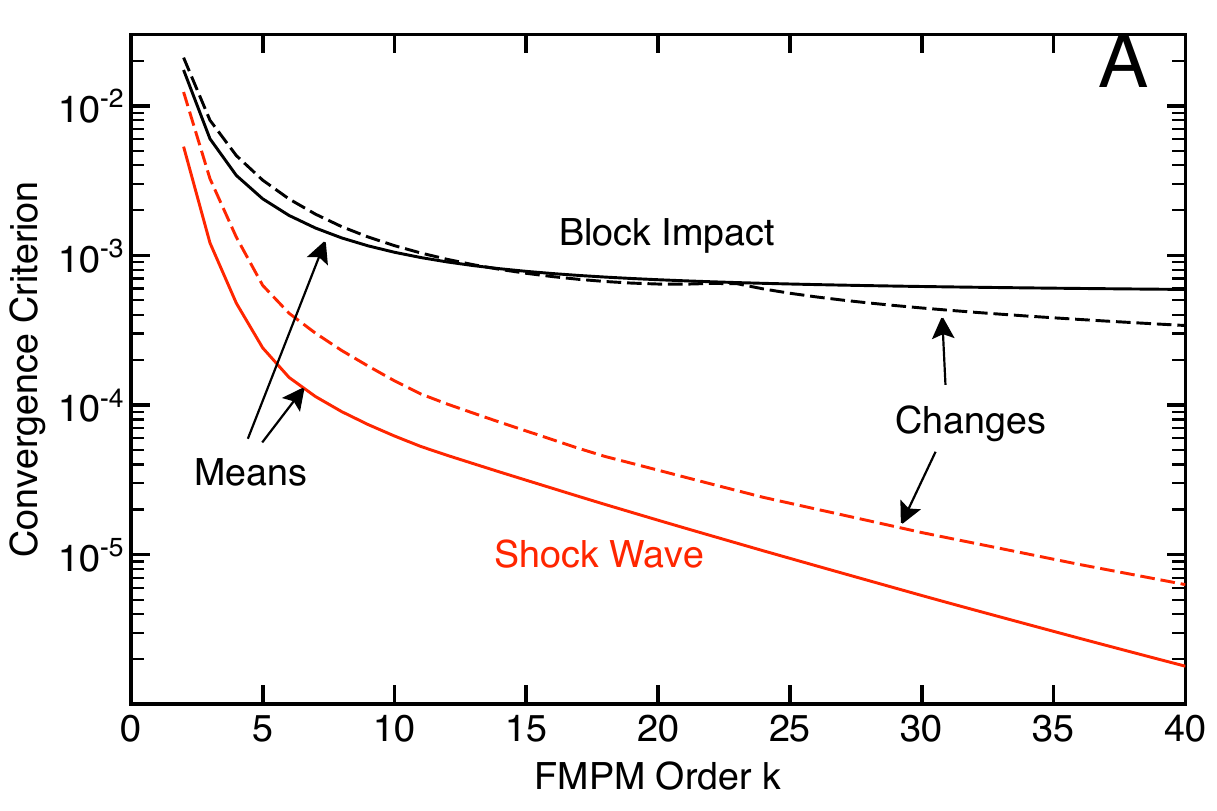}
\hspace{0.02\linewidth}
\includegraphics[width=0.45\linewidth]{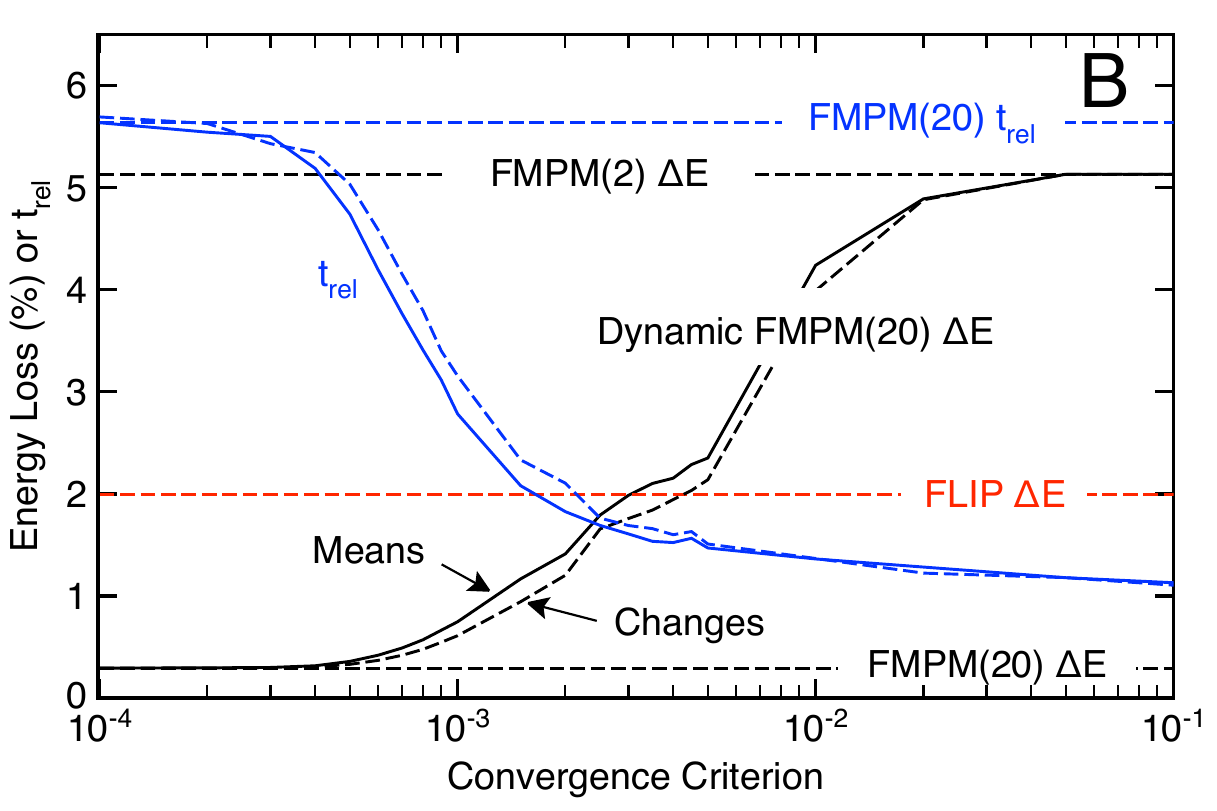}
\caption{A. Sample convergence of ``Means'' and ``Changes'' metrics for time step in the middle of the block impact or shock wave simulations. B Energy loss and calculation time relative to FLIP calculations as a function of critical metric to find FMPM($k$) convergences for block impact simulations. Solid lines used ``Means''  metric. Dashed lines used ``Changes'' metric. Horizontal lines plot results for non-dynamic FMPM(2), FMPM(20), or FLIP.}
\label{dynamic}
\end{figure}

Figure~\ref{dynamic}A plots FMPM($k$) convergence metrics for an arbitrary time step in the middle of a two-block impact simulation. The calculations are converging, but after an initial rapid drop, the convergence is slow. The ``Means'' criterion appears to level off rather than approach zero, which might indicate results oscillating around an ideal solution. The ``Changes'' metric is slightly better, but also converges slowly. 

Figure~\ref{dynamic}B gives dynamic FMPM(20) results for the two-block impact problem plotting energy dissipation for dynamic calculations (black curves) and simulation times relative to FLIP calculations (blue curves) as a function of convergence limit. The solid curves used the ``Means'' metric while the dashed curves used the ``Changes'' metric. When the convergence limit is high ($\gtrsim0.1$), the loop always converges with FMPM(2) and the energy dissipation and simulation time matches the results for non-dynamic FMPM(2) $\Delta E$ (dashed line) and matches the FMPM(2) relative time. When the convergence limit is low ($\lesssim0.0001$), the FMPM loop never converges (except for the initial time steps before impact) and both energy dissipation and relative simulation time match the results for non-dynamic FMPM(20). The net effect is a simulation that simply uses FMPM(20) on every time step after the initial impact. Between these limits, dynamic FMPM(20) has some benefits. Using the ``Means'' metric, energy dissipation is reduced to matching FLIP with simulation time only $1.6\times$ longer and reaches half the FLIP dissipation with simulation time only $2.4\times$ longer. Results with the ``Changes'' metric were the same to match FLIP energy dissipation but had a slight improvement to reach half the FLIP dissipation with simulation time only $2.3\times$ longer.

While dynamic FMPM($k$) is easy to implement, the results on this one problem were mixed. A carefully chosen criterion provided some benefit, but choosing the criterion too high reverts to FMPM(2) and too low reverts to maximum provided order with no benefit of dynamic calculations.  Getting dynamic FMPM(20) to half the FLIP dissipation required a precise value for convergence limit (0.0015) and that value was within the region were results changed rapidly.

\begin{figure}
\centering
\includegraphics[width=0.45\linewidth]{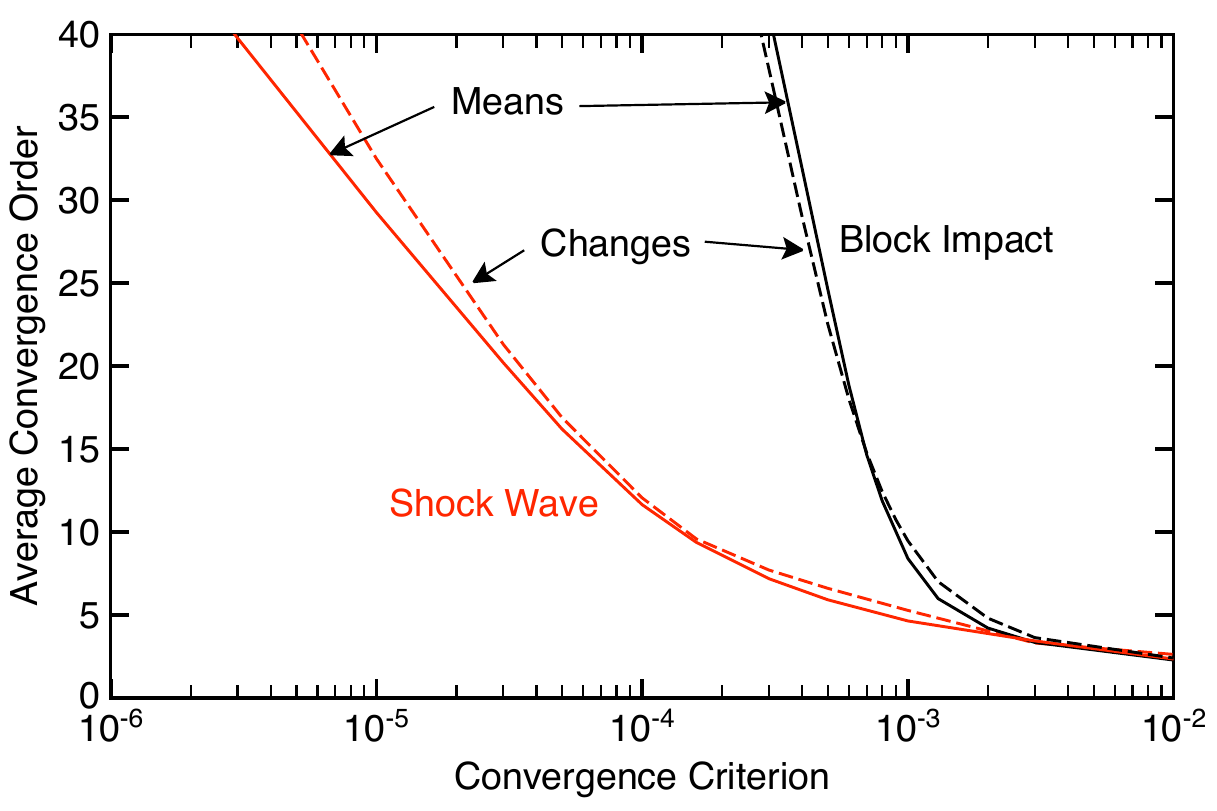}
\caption{The average FMPM($k$) order for convergence as a function of the convergence criterion for the two-block impact and shock wave examples. Solid lines are when using the ``Means'' metric while dashed lines are when using the ``Changes'' metric.}
\label{dynamicorder}
\end{figure}

This two-block impact might be particularly challenging. The impact event induces dynamic stress waves that continue forever in elastic modeling. Perhaps such waves always need high order for accurate calculations? For problems where stress waves either dissipate or approach some steady state conditions, dynamic FMPM might provide more benefits. For example, the shock wave simulations in \Fig{ShockStick} noted that FMPM($k$) results for $k>4$ were identical to FMPM(4) results. In this problem, the simulations starts by developing a shock wave front, but than that front translates though the object with minimal spatial variations (besides the translations). For another test of dyamic FMPM($k$), the method was tried on the shock wave example. Figure \ref{dynamic}A shows convergence of the two metrics for an arbitrary time step in the middle of a simulation. This response is more promising. The metric spans a wider range in values and continues to approach zero as order $k$ increases. Figure \ref{dynamicorder} compares the average FMPM($k$) order required to reach convergence as a function of convergence criterion for both impacting blocks and a shock wave. The block impact order increases very rapidly. In contrast, shock wave order increases more slowly over a wide range of convergence metrics. Nevertheless, the shock wave calculations still needed very high order when the convergence criterion got small. It needed high order despite the observation that using $k>4$ provides little or no change in overall simulation results.

\vskip6pt
\noindent \emph{Remark}: Overall, these first attempts at dynamic FMPM($k$) provided limited benefit. The main problem is that choosing the wrong convergence criterion can cause the FMPM($k$) calculations to require very high order even when such high orders provide little benefit to simulation results. Rather then rely on dynamic FMPM($k$) to control simulation order, a simpler practice is to use modest order. Experience suggests FMPM($k$) benefits have diminishing beneftis for $k>5$ suggesting that FMPM(4) (or slightly higher $k$) without implement dynamics FMPM($k$) is likely the most efficient approach. Alternatively, one could use dyamic FMPM($k$) but choose a modest value, such as $k=6$, as the maximum allowed order. This situation could change if future work devises new convergence metrics that better recognize diminishing benefits of continuing to higher order. 

\subsection{Other FMPM Calculations}

Besides above details on revising FMPM($k$) methods, other sections of MPM modeling are affected by these changes:

\begin{itemize}

\item Transport Modeling: The original FMPM($k$) methods in Ref.~\cite{Nairn:198} were recently adapted for solving transport equations by MPM \cite{Nairn:2024}. Instead of a full mass matrix, it used FMPM($k$) methods to approximate full transport capacity matrix inverse. The methods in Ref.~\cite{Nairn:2024} are easily converted to using the revised methods in this paper. In fact, the implementation of this paper's revised methods in {\tt OSParticulas} \cite{Nairn:Other2} and {\tt NairnMPM} \cite{Nairn:Git} created a generic FMPM loop task with calculations for mass matrix inverse or transport capacity matrix as subclasses to the generic parent class. By this approach, implementing revised FMPM($k$) to mechanics modeling automatically implements revised FMPM($k$) methods for transport modeling.

\item Explicit Cracks: Virtually all methods for modeling explicit cracks in MPM are unaffected by whether the modeling is using standard MPM or FMPM($k$). The one exception is modeling of crack contact. Those contact calculations are very similar to multimaterial MPM contact except that normal to the contact surface can be determined from explicit crack surfaces instead of logistic regression calculations. Fortunately, the methods derived above  that allow FMPM($k$) to work with lumped mass contact derived in Section~\ref{MMContact} are directly translatable to crack contact calculations.

\item Imperfect Interfaces: MPM can model imperfect interface using either explicit cracks \cite{Nairn:122} or as a modification of contact calculations between materials \cite{Nairn:146}. In brief, these methods calculate force needed to implement an interface law that models force as a function of interfacial discontinuity. Imperfect interfaces are  best implemented within MPM contact mechanics calculations \cite{Nairn:146}, but they use different physical models for calculating momentum changes applied to nodes. To be compatible with FMPM($k$), imperfect interface calculations would need to implement incremental calculations analogous to contact methods described here. Those required changes will be addressed in a future publication along with improvements over prior imperfect interface methods \cite{Nairn:2026aa}.

\end{itemize}

\section{Conclusions}

The revised FMPM($k$) methods in this paper are identical to original FMPM($k$) calculations in Ref.~\cite{Nairn:198}, but the revised FMPM loop has three advantages. First, it is easier to implement (it eliminates some $k$- and $\ell$-dependent scaling factors). Second, it provided a path to resolving all conflicts between FMPM($k$) and various MPM features such a grid-based velocity boundary conditions, multimaterial and crack contact, and imperfect interfaces \cite{Nairn:2026aa}. Third, physical interpretation of $\Delta \vec v(\ell) = \vec v^+(\ell) - \vec v^+(\ell-1)$ being change in velocity when using FMPM($\ell$) instead of FMPM($\ell-1$) that is independent of $k$ made it easy to implement new options such as blending with FMPM(2) mass matrix or exploring dynamic convergence of simulation results.

All modification derived here were tested to high order, which now enables use of FMPM($k$) to any order required for good simulation results. In practice, FMPM(2) alone greatly reduces the energy dissipation seen in PIC methods or FMPM(1). Unfortunately, FMPM(2) may not reduce it enough for optimal results. The solution is for MPM codes to implement FMPM($k$) of any order. Practical experience suggest that FMPM(4) or FMPM(5) is often enough such that moving to higher order has diminishing returns. Those tempted to implement only FMPM(2), should realize that once that is available, implementing FMPM($k$) requires virtually no addition coding. A complete implementation of FMPM(2) still requires all the coding in Table~\ref{FMPMcode} except that the loop is only done once. Notably, it still needs to add velocity boundary condition and contact calculations as indicated in lines (1) and (2) in Table~\ref{FMPMcode}. Once FMPM(2) is implemented, the only change needed to implement FMPM($k$) is the enclose FMPM(2) calculations in a loop that is repeated $k-1$ times. The full FMPM($k$) methods in this paper are available in {\tt OSParticulas} code \cite{Nairn:Other2} and publicly available in {\tt NairnMPM} version 19 or newer \cite{Nairn:Git}.

\bibliography{ReferencesFMPM}{}

\end{document}